\def\msol{\rm{M}$_{\odot}$}
\def\lsol{\rm{L}$_{\odot}$}

\def\HII{H\,{\sc ii}}
\def\HI{H\,{\sc i}}
\def\arcsec{$^{\prime}$$^{\prime}$}
\def\arcmin{$^{\prime}$}
\def\deg{$^{\circ}$}
\def\micron{\,$\mu$m}
\def\etal{\textit{et al.}}

\documentclass{aa}

\usepackage{longtable,lscape,graphicx,float,txfonts,natbib,multirow,color,ulem} 

\begin{document}

\title{The RMS Survey: The Bolometric Fluxes and Luminosity Distributions of Young Massive Stars.\thanks{Tables \ref{T:results_full} and \ref{T:nofir_results_full} are only available in electronic form at the CDS via anonymous ftp to cdsarc.u-strasbg.fr (130.79.125.5) or via http://cdsweb.u-strasbg.fr/cgi-bin/qcat?J/A+A/.}}

\author{J.~C.~Mottram\inst{1,2}~\thanks{E-mail:joe@astro.ex.ac.uk}
  \and
  ~M.~G.~Hoare\inst{1}
  \and
  ~J.~S.~Urquhart\inst{3}
  \and
  ~S.~L.~Lumsden\inst{1}
  \and
  ~R.~D.~Oudmaijer\inst{1}
  \and
  ~T.~P.~Robitaille\inst{4}
  \and
  ~T.~J.~T.~Moore\inst{5}
  \and
  ~B.~Davies\inst{6,1}
  \and
  ~J.~Stead\inst{1}
}

\institute{School of Physics and Astronomy, University of Leeds, Leeds, LS2 9JT, UK
  \and
Department of Physics and Astronomy, University of Exeter, Exeter, Devon, EX4 4QL, UK  
  \and
Australia Telescope National Facility, CSIRO Astronomy and Space Science, Sydney, NSW 2052, Australia
 \and
Harvard-Smithsonian Center for Astrophysics, 60 Garden Street, Cambridge, MA 02138, USA
 \and
Astrophysics Research Institute, Liverpool John Moores University, Twelve Quays House, Egerton Wharf, Birkenhead, CH41 1LD, UK
\and
Rochester Institute of Technology, 54 Lomb Memorial Drive, Rochester, NY 14623, USA
}

\date{Received XXX 2010 / Accepted XXX 2010}

\abstract
   {The Red MSX Source (RMS) survey is returning a large sample of massive young stellar objects (MYSOs) and ultra-compact (UC) \HII{} regions using follow-up observations of colour-selected candidates from the MSX point source catalogue.}
   {To obtain the bolometric fluxes and, using kinematic distance information, the luminosities for young RMS sources with far-infrared fluxes.}
   {We use a model spectral energy distribution (SED) fitter to obtain the bolometric flux for our sources, given flux data from our work and the literature. The inputs to the model fitter were optimised by a series of investigations designed to reveal the effect varying these inputs had on the resulting bolometric flux. Kinematic distances derived from molecular line observations were then used to calculate the luminosity of each source.}
   {Bolometric fluxes are obtained for 1173 young RMS sources, of which 1069 have uniquely constrained kinematic distances and good SED fits. A comparison of the bolometric fluxes obtained using SED fitting with trapezium rule integration and two component greybody fits was also undertaken, and showed that both produce considerable scatter compared to the method used here.}
   {The bolometric flux results allowed us to obtain the luminosity distributions of YSOs and UC\HII{} regions in the RMS sample, which we find to be different. We also find that there are few MYSOs with L~$\geq$~10$^{5}$\lsol{}, despite finding many MYSOs with 10$^{4}$\lsol{}~$\geq$~L~$\geq$~10$^{5}$\lsol{}.}


\authorrunning{J.~C.~Mottram et al.}
\titlerunning{The Bolometric Fluxes of Young Massive Stars}
\maketitle

\section{Introduction}
\label{S:intro}

Though they account for a tiny fraction of the galactic stellar population, massive stars (M~$\geq$~8~\msol{}) dominate the galaxies within which they form and evolve. The radiation and outflows from young massive stars act as feedback processes, altering their parent molecular cloud and thus influencing current and subsequent generations of nearby star formation. Eventually, the ionising fronts of \HII{} regions and flows of material driven by stellar winds from these stars can lead to the destruction of the molecular cloud itself, ending further star formation altogether. However, our understanding of these effects is still in its infancy, with further study limited by the lack of large well-selected samples from which to derive the general properties of young massive stars and to study how these evolve in terms of both stellar mass and age.

To this end, the Red MSX Source (RMS) survey \citep{Hoare2005,Urquhart2008b} has carried out a series of follow-up observations of a sample of candidate massive young stellar objects (MYSOs) which was colour-selected from the MSX point source catalogue \citep[][]{Egan1999,Egan2003a} by \citet{Lumsden2002}. We define an MYSO as a radio-quiet mid-infrared (IR) point source which has reached a luminosity close to its final value, but where  accretion is probably still ongoing and an ionised \HII{} region has yet to form. The mid-IR colours of MYSOs are similar to those of both young and evolved dust-enshrouded objects (such as ultra-compact (UC)\HII{} regions, low-mass YSOs, evolved stars, proto-planetary nebulae and planetary nebulae (PNe)). Therefore additional observations are required to identify contaminants and gain information about the candidates. The RMS survey has undertaken $\sim$~1\arcsec{} radio continuum observations to identify radio loud UC\HII{} regions and PNe \citep{Urquhart2007a,Urquhart2009}, $\sim$~1\arcsec{} mid-IR imaging outside the GLIMPSE survey region \citep{Benjamin2003,Churchwell2009} to search for candidates which show multiple and/or extended sources at higher resolution \citep[][]{Mottram2007} and near-IR low-resolution spectroscopy in order to identify evolved stars and weak, unresolved UC\HII{} regions \citep[e.g.][]{Clarke2006}. We have also undertaken pointed $^{13}$CO J=1$-$0 or J=2$-$1 observations to obtain kinematic velocities of our sources in order to calculate kinematic distances \citep{Urquhart2007c,Urquhart2008a}.

In a previous paper \citep{Mottram2010} we presented new measurements of the fluxes of RMS sources in the far-IR (by which we mean 30\micron{} to 300\micron{}) using IRAS Galaxy Atlas \citep[IGA,][]{Cao1997} and Spitzer MIPSGAL \citep{Carey2009} images. These are important if the luminosities of RMS sources are to be obtained, as the spectral energy distribution (SED) of young massive stars in both the MYSO and UC\HII{} region phases peak between 60\micron{} and 120\micron{}. 

In this paper we will use the far-IR data from \citet{Mottram2010} along with other data from RMS observations and the literature to obtain the SEDs of these sources (\S\ref{S:data}). In \S\ref{S:tests} we discuss tests of the SED model fitter of \citet{Robitaille2007a}, which are used to inform the choice of input parameters for the fitter and evaluate what represents a good result. The results of the SED fitting and derived bolometric fluxes are presented in \S\ref{S:results}, along with the calculation of filter flux ratios and flux estimates for those sources without far-IR fluxes derived from these ratios. In \S\ref{S:discussion} we present a comparison with other methods for obtaining the bolometric flux from SEDs, then discuss the luminosity distributions of RMS sources in terms of their source type, before we reach our conclusions in \S\ref{S:conclusions}.

\section{Input data \& the model fitter}
\label{S:data}

The SEDs of the candidate MYSOs presented in this paper are primarily based on fluxes from v2.3 of the MSX point source catalogue \citep[PSC,][]{Egan2003a} used to define the RMS sample \citep[see][]{Lumsden2002} and far-IR fluxes from \citet[][]{Mottram2010}. Far-IR fluxes obtained from 18\arcsec{} resolution MIPSGAL \citep[][]{Carey2009} 70\micron{} data were used where available in preference to those from 1\arcmin{}~$\times$~1.7\arcmin{} IRAS Galaxy Atlas \citep[IGA,][]{Cao1997} data. Sources without full detections in either far-IR data are not included, as there is therefore no data to constrain the far-IR peak of the SED for these objects (the result of fitting the SED without far-IR data will be discussed in \S\ref{S:tests_data}). While the original coordinates for all sources were those from the MSX PSC, in some cases these have been updated to those from higher-resolution data.

The subsections below discuss additional data included, where possible, to better constrain the SEDs (\S\ref{S:data_timmi2} $-$ \S\ref{S:data_submm_and_mm}), and provide distance information (\S\ref{S:data_distances}).

\subsection{Mid-IR imaging}
\label{S:data_timmi2}

In order to supplement these data, we also include fluxes from our $\sim$~1\arcsec{} resolution 10.4\micron{} TIMMI2 observations \citep[][]{Mottram2007}. The combined 3$\sigma$ uncertainty in the offset between MSX and TIMMI2 positions for all sources observed with TIMMI2 was found to be 7.6\arcsec{}. Therefore, where TIMMI2 observations show no source within a radius of 7.6\arcsec{} of the source coordinates, a non-detection upper limit flux is used based on the uncertainty in the 10.4\micron{} TIMMI2 flux.

From TIMMI2 and other high-resolution mid-IR imaging, it is clear that some of the flux detected within the MSX beam (18\arcsec{}) in star formation regions is due to diffuse emission not directly relating to the source of interest. Therefore, the bolometric flux of larger-beam observations should be apportioned based on the higher-resolution data. In these cases, the apportion ratio was therefore calculated as the ratio of the 10.4\micron{} flux to the MSX 8\micron{} flux. Where the TIMMI2 flux is larger than the MSX 8\micron{} flux, the ratio is set to 1. The individual fluxes from large-beam observations (i.e. MSX, MIPSGAL, IGA, sub-mm and mm) were then multiplied by the apportion ratio before input into the model fitter and the uncertainties of these fluxes calculated as the addition in quadrature of the uncertainty in the apportion ratio and the measurement uncertainties. 

In using a simple ratio of the 10.4\micron{} flux to the MSX 8\micron{} flux we assume that all the sources detected in the TIMMI2 observations have the same SED shape and that the 9.7\micron{} silicate feature affects both fluxes equally. While this is an obvious simplification, it is certainly an improvement over not performing this correction and the absence of higher resolution data at longer wavelengths for all our sources makes a more detailed determination impossible at this time. This method was required for 178 out of a total of 1183 sources.

Similarly, some RMS sources have detections at 8\micron{} in the GLIMPSE PSC, in which case this flux is used to calculate the apportion ratio as for TIMMI2 sources. 182 out of a total of 1183 sources were apportioned in this way. In addition, in some cases it is clear from higher-resolution mid-IR imaging that more than one YSO or \HII{} region contributes significantly to the flux within the MSX beam. For these sources, the individual components are given letter designations and apportioning carried out where possible. Where we do not have measured fluxes, such as cases which can be seen in the GLIMPSE 8\micron{} images but do not have 8\micron{} fluxes in the PSC, it is still clear that each individual source does not contain the total MSX flux. Therefore the remaining fraction of the total flux of the larger-beam observations is split equally between those sources which cannot be apportioned using TIMMI2 or GLIMPSE 8\micron{} fluxes. This is performed for 136 out of a total of 1183 sources.

\subsection{2MASS}
\label{S:data_2mass}

In order to help constrain the near-infrared side of the SED, $J$, $H$ and $K_{s}$ band fluxes from the nearest 2MASS PSC \citep[][]{Skrutskie2006} entry within 7.6\arcsec{} (see above) are used. Though this radius may seem large, it is still less than half the full-width half maximum (FWHM) of MSX (18\arcsec{}) and is motivated by the uncertainty in the source coordinates. In the case of sources with TIMMI2 observations, the 2MASS counterparts identified by \citet{Mottram2007} are used instead. Where there is no 2MASS detection within 7.6\arcsec{} of the MSX PSC coordinates, or where inspection of the 2MASS images during object classification indicated that the nearest source is unlikely to be the near-IR counterpart to the MSX target, the noise in each band is calculated in the following way in order to estimate an upper limit flux. All 2MASS PSC entries which are detected in other bands but not in the band in question within 1\arcmin{} of the target coordinates are identified. An average of the upper limit fluxes in the band in question of all such sources is then calculated. While some 2MASS counterparts may be mis-identified by this method, the SEDs of young massive sources are dominated by emission in the far-IR, so this is unlikely to have an overriding influence on the calculated bolometric flux.

\subsection{Sub-millimetre and Millimetre Continuum}
\label{S:data_submm_and_mm}

SCUBA 450\micron{} and 850\micron{} data from the legacy archive \citep{DiFrancesco2008} are included, where available, for those sources with an offset between the SCUBA observations and the source coordinates $\leq$~14\arcsec{}, the beam full-width at half maximum (FWHM) of SCUBA at 850\micron{}. The FWHM is used rather than the pointing error (usually of order $\pm$~5\arcsec{}) as many of these SCUBA sources associated with RMS young massive stars are larger than the beam size. In cases where the offset was larger than 14\arcsec{} but smaller than the source size at 850\micron{}, the SCUBA fluxes were used as upper limits to the true flux at these wavelengths. The percentage errors on the SCUBA 450\micron{} and 850\micron{} fluxes are 50~$\%$ and 20~$\%$ respectively \citep{DiFrancesco2008}.

Millimetre continuum observations at 1.2~mm with the SEST imaging Bolometer array (SIMBA) on the Swedish ESO submillimetre telescope (SEST) by \citet{Faundez2004}, \citet{Hill2005} and \citet{Beltran2006}, along with 1.2~mm observations with the Max-Planck Millimetre Bolometer (MAMBO) on the Institut de Radio Astronomie Millimetrique (IRAM) 30~m telescope by \citet{Beuther2002a} were also included. Originally, all sources with offsets from the source coordinates less than or equal to the beam size (24\arcsec{} for SIMBA and 11\arcsec{} for MAMBO) were included as detections and those with offsets larger than the beam but smaller than the source size as upper limits based on the source flux. However, in some cases the source sizes as derived by Gaussian fits are large compared to the beam size and$/$or the source is irregular and extended at 1.2~mm, resulting in a much higher flux at this wavelength than is consistent with the rest of the data. All 1.2~mm images were therefore examined and compared to MSX images in order to determine which observations should be regarded as upper limits to the true source flux.

The uncertainties on detected sources are taken as the rms noise from the various data sets, i.e. 40~mJy \citet{Faundez2004}, 40~mJy for \citet{Beltran2006} and 15~mJy for \citet{Beuther2002a}. The observations of \citet{Hill2005} were observed over multiple runs, so in order to be conservative the largest rms noise value from these runs (i.e. 150~mJy) was used for all data. For non-detections, or sources with offsets from the MSX position larger than the measured size of the detection at 1.2~mm, 3~$\sigma$ of the appropriate minimum background flux levels are used as an upper limit to the 1.2~mm source flux.

\subsection{Kinematic Distances}
\label{S:data_distances}

Kinematic distances have been obtained for the majority of RMS sources, calculated from kinematic velocities using the \citet{Brand1993} rotation curve, assuming the galactocentric distance and velocity of the solar system to be 8.5~kpc and 220~kms$^{-1}$ respectively \citep{Urquhart2007c,Urquhart2008a}. All sources with a galactocentric radius $<$~8.5~kpc have a kinematic distance ambiguity caused by the fact that two points along the line of sight, one near and one far, have the same line of sight velocity in the galactic rotation curve, though at the tangent point these are the same and so not an issue. We have solved this distance ambiguity in the northern hemisphere \citep{Urquhart2010} by cross-matching our sources with a sample of clouds identified by \citet{Rathborne2009} in the $^{13}$CO Boston University Five College Radio Astronomy Observatory (BU-FCRAO) Galactic Ring Survey \citep[GRS, 18\deg{}~$\leq$~l~$\leq$~55.7\deg{}, $\mid$b$\mid$~$\leq$~1\deg{};][]{Jackson2006} for which the ambiguity has already been solved by \citet{RomanDuval2009}. In the southern hemisphere we are using data from the Southern Galactic Plane Survey \citep[SGPS,][]{McClure-Griffiths2005} and the \HI{} self-absorption method \citep[][]{Liszt1981} to solve the ambiguity to our sources (Urquhart \etal{}, 2010b, in prep.).

For sources where there are unsolved near-far distance ambiguities, fits are calculated for all distances. These 33 sources are included in the Table of results (table~\ref{T:results_full}) presented in \S\ref{S:results} as the average of these fits, but are not included in the analysis.

\section{Fitting the spectral energy distributions of young massive stars}
\label{S:tests}

This section discusses how the bolometric flux is obtained from model fits to spectral energy distributions (\S\ref{S:tests_fitter}), then describes a test designed to investigate how use of an incorrect distance in the fitting process affects the accuracy of the resulting bolometric flux (\S\ref{S:tests_distance}). The choice of the $A_{V}$ range that the fitter is allowed is discussed (\S\ref{S:tests_extinction}), followed by a study of the relative importance of the various data sets that make up the SED data (\S\ref{S:tests_data}). The investigations in \S\ref{S:tests_distance} and \S\ref{S:tests_data} were run using the data for G034.7569+00.0247 as the data set for this probable MYSO is one of the most complete.

\subsection{The model SED fitter}
\label{S:tests_fitter}

The radiative transfer models of \citet{Robitaille2006} cover a range of different stellar, disk and envelope properties at a range of angles of inclination. The model fitter of \citet{Robitaille2007a} then takes these models, convolves them with a user-specified set of filters, allows a range in distance and $A_{V}$ and then minimises the $\chi^{2}$ between the model filter fluxes and the input data. We use this model fitter in preference to other methods as it takes into account dust features, such as the 9.7\micron{} silicate feature, and consider a more realistic geometry for the circumstellar material than a simple sphere. It is important to mention that a good fit to the data does not prove that the geometry or conditions present in the model are a good description of the observed source, rather that the model is consistent with the observational data. In particular, it may be the case that the data are insufficient to fully constrain the 14 parameters used to construct the models, so it is not appropriate to infer knowledge of variables that the SED data alone do not specifically constrain \citep[][]{Robitaille2008}. \citet{deWit2010} found that the best fitting models returned by the fitter for MYSOs usually have quite massive dust disks which are ruled out by mid-IR interferometric observations.  In addition, due to the non-random nature in which the parameters were sampled care must be taken to establish that observed trends in the data are not a result of trends in the parameter space \citep[see][for a full discussion]{Robitaille2008}.

The bolometric flux of the fit is, however, well constrained by the fits, as this relates directly to the SED \citep{Robitaille2008}. The following analysis will therefore focus on this property, while comparisons between the bolometric flux obtained by SED fits with other methods will be addressed in \S\ref{S:discussion_other}.

It is important that values derived from the results of the model fitter include several models, in order to obtain a reasonable estimate of whether these results are constrained and unique with the data used. Therefore the number of fits within a $\Delta\chi^{2}$=2 of the best fit was calculated \citep[we use the same definition of $\chi^{2}$ as][]{Robitaille2007a}. If there were less than 10 fits included within $\chi_{best}^{2}$~+~$\Delta\chi^{2}$ then $\Delta\chi^{2}$ was increased by integer steps until this was no longer the case. This method was used rather than simply finding the best 10 fits so that the spread in fits was not artificially limited if many models are good fits to the data with a range of values for distance, bolometric flux and $A_{V}$. A minimum of 10 fits were required in order to ensure that one or two fits did not strongly bias the results.

\begin{figure*}
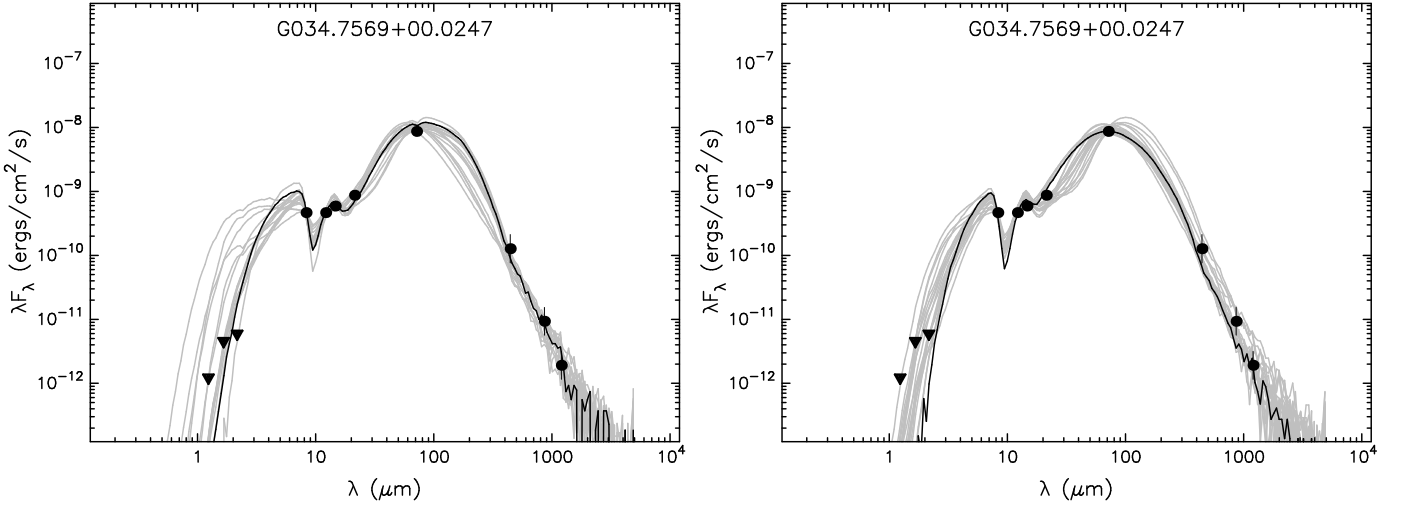

\center
\includegraphics[width=0.49\textwidth]{G034.7569+00.0247_13.eps}
\includegraphics[width=0.49\textwidth]{G034.7569+00.0247_43.eps}
\caption{Example SED fits with allowed distances of 2.0~$\pm$~1.0~kpc (left) and 5.0~$\pm$~1.0~kpc (right) respectively. The circular points indicate flux detections while the downward pointing triangles indicate upper limit fluxes. The solid line shows the best fit to the data, while the grey lines indicate other fits within a given $\Delta\chi^{2}$ of the best fit.}
\label{F:tests_distance_seds}
\end{figure*}

\begin{figure*}
\center
\includegraphics[width=0.49\textwidth]{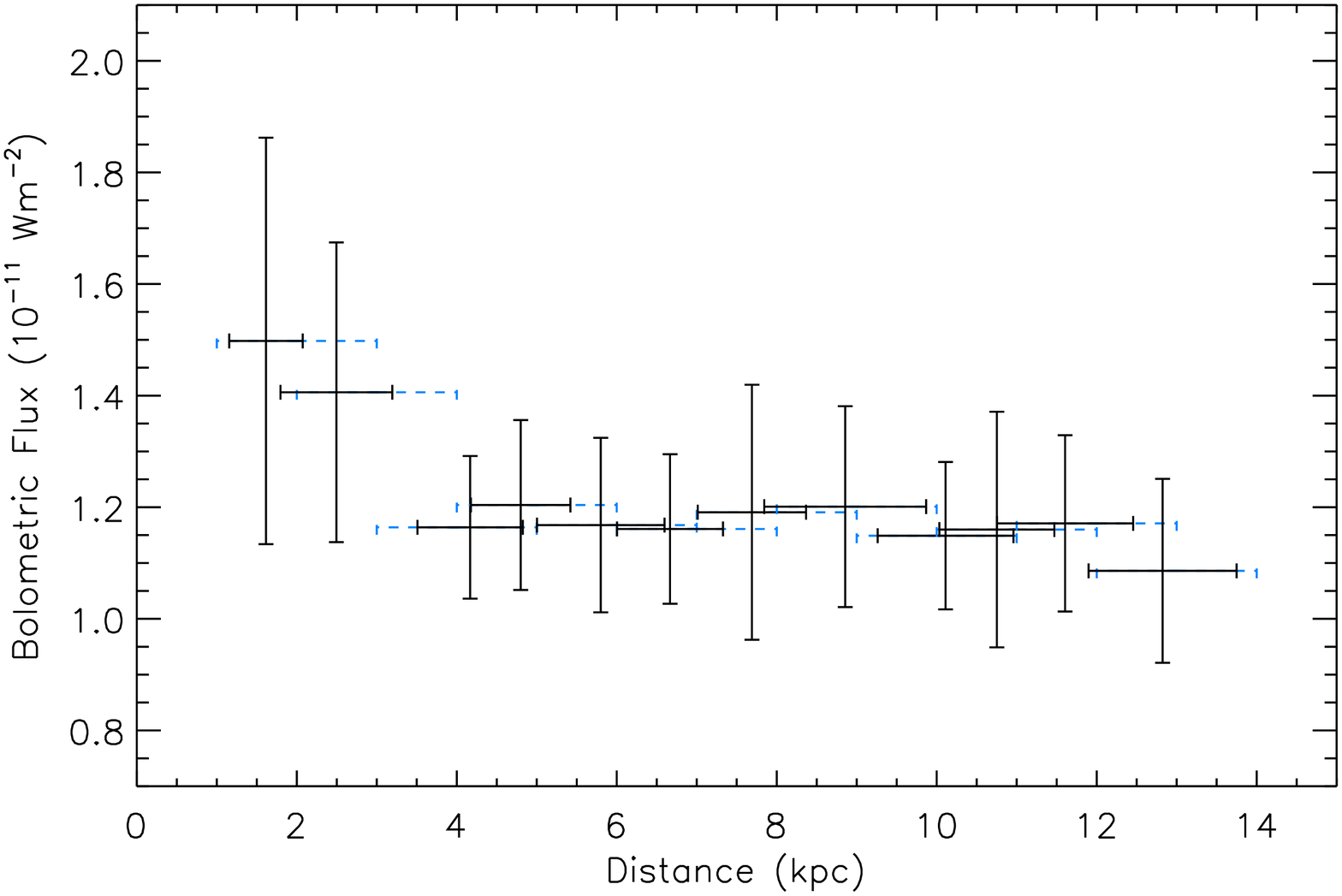}
\includegraphics[width=0.49\textwidth]{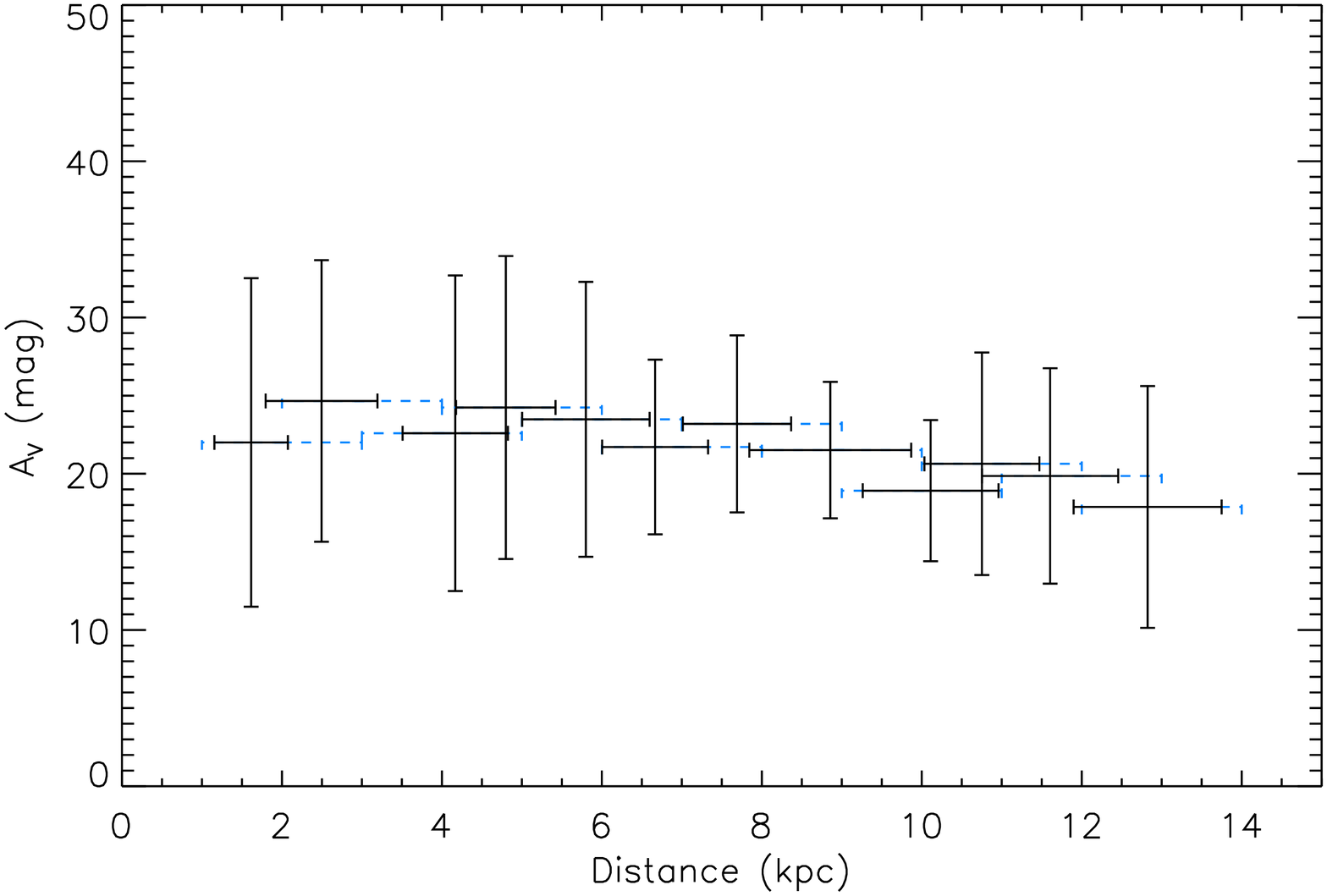}
\caption{Bolometric flux (left) and $A_{V}$ (right) vs. the mean model distance obtained from SED fits with specified model distances d~$\pm$~1.0~kpc with d varying between 2~kpc and 13~kpc for the data of G034.7569+00.0247. The black horizontal error bars indicate the uncertainty on the mean model distance while the dashed horizontal error bars indicate the range of allowed model distances.}
\label{F:tests_distance_allruns}
\end{figure*}

The model bolometric flux for the source was obtained by using the model luminosity and model distance derived for each individual fit to calculate the bolometric flux for each fit ($f_{i}$). The luminosity returned by the model fitter is the combination of that for the central source, accretion onto the central source and accretion in the disk, which is then corrected for foreground extinction consistent with the SED, and is therefore independent of viewing angle unlike the SED itself. Next the weighted mean and weighted standard deviation were calculated of all fits within $\chi_{best}^{2}$~+~$\Delta\chi^{2}$ to give the flux and flux uncertainty using:

\begin{equation}
\bar{F}_{w}~=~\frac{\sum_{i=1}^{N}~w_{i}~f_{i}}{\sum_{i=1}^{N}~w_{i}}, \delta\bar{F}_{w}~=~\sqrt{\frac{N^{'}}{(N^{'}-1)}~\frac{\sum_{i=1}^{N}~w_{i}(f_{i}~-~\bar{F}_{w})^{2}}{\sum_{i=1}^{N}~w_{i}}}
\label{E:tests_flux}
\end{equation}

\noindent
where $N^{'}$ is the number of non-zero weights, and the weights are calculated from the $\chi^{2}$ as:

\begin{equation}
w_{i}~=~e^{-(\chi^{2}_{i}/2)}
\label{E:tests_weights}
\end{equation}

\noindent
The $\chi^{2}$ values are used to weight the fits so that the best model fitted to the data has more impact on the results than poorer fits, but that the results are not heavily skewed away from the mean. This method is also used to calculate the mean model distance, the mean $A_{V}$, and mean filter flux ratios given the results for individual model fits to the data.

\subsection{Model Distance}
\label{S:tests_distance}

A distance range is required by the model fitter within which to fit the observed data. Therefore the test described below was undertaken in order to explore the distance dependence of the bolometric flux obtained from the model fitter, and thus what effect an incorrect distance has on the model derived flux.

Fits were performed to the data for G034.7569+00.0247 for model distances $d$~$\pm$~$\delta$$d$~kpc, with $d$ varying between 2~kpc and 13~kpc in 1~kpc integer steps. Example SED fits for 2.0~$\pm$~1.0~kpc and 5.0~$\pm$~1.0~kpc \citep[the kinematic distance for G034.7569+00.0247 is 5.0~$\pm$~0.7~kpc,][]{Urquhart2008a} are shown in Figure~\ref{F:tests_distance_seds} (left and right respectively). For all runs of the model fitter, $\delta$$d$ is set to 1.0~kpc as this is the order of the typical error in the kinematic distance for RMS sources.

The model fitter divides the user-specified distance range logarithmically in order to obtain model distances after which the model SED fluxes are convolved with common filter bands and interpolated to the apertures used to obtain the data. The logarithmic binning was set to 0.01 for this test, while 0.025 was used for the other tests and the results presented in \S\ref{S:results}. The results were checked for a few sources at the lower model distance binning step and the differences were found to be minimal. At small model distances the apertures may not encompass the whole model but only the inner region, the calculation of which can require large amounts of computer memory. Therefore, where the distance to the source was less than 2.4~kpc, the distance given to the fitter for the main results was reset to this minimum limit.

The results of the fits where $d$ is varied for the data of G034.7569+00.0247 are shown in Figure~\ref{F:tests_distance_allruns}, in terms of bolometric flux (left) and $A_{V}$ (right) vs. the mean model distance as calculated using the weighted mean and weighted standard deviation of the model distances for the individual fits within $\Delta$$\chi^{2}$ of $\chi^{2}_{best}$. The $A_{V}$ is relatively constant with distance for these fits (see Figure~\ref{F:tests_distance_allruns}), suggesting it is constrained by the data. The increase in the bolometric flux and error at small distances is due to the source being so close that the $A_{V}$ is insufficient to stop emission in the visual being observed (see Figure~\ref{F:tests_distance_seds}). The model fitter allows this as the near-IR fluxes are upper limits, and so are treated as weaker constraints on the SED than the detections at longer wavelengths. Overall, though the bolometric flux appears to decrease with increasing distance, the change is of similar order or smaller than the uncertainties in the fluxes, and so the results can be considered distance independent.

\subsection{Visual Extinction}
\label{S:tests_extinction}

The visual extinction ($A_{V}$) between Earth and the edge of the circumstellar material considered by the model is an important free parameter in fitting model SEDs to observed data. Restricting this value too much could result in poor fits, particularly between the near/mid-IR and far-IR/sub-mm parts of the SED, while allowing too much freedom could lead to unphysical solutions. The extinction law model assumed was calculated by \citet{Robitaille2007a} using the method of \citet{Kim1994} for a typical galactic ISM, taking into account the results of \citet{Indebetouw2005} with regards to mid-IR extinction.

\begin{figure}
\center
\includegraphics[width=0.49\textwidth]{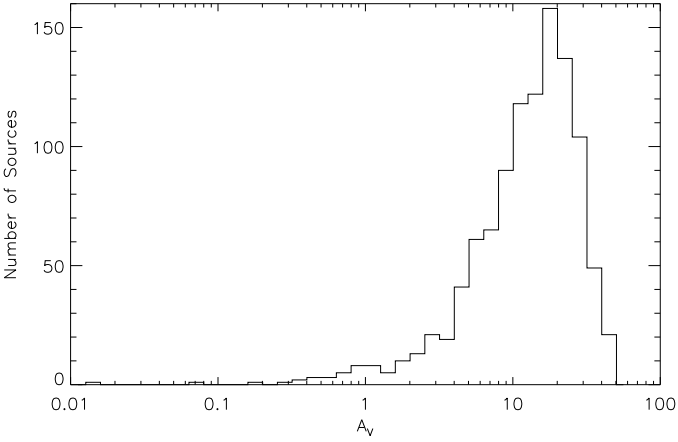}
\caption{Left: A histogram of $A_{V}$ derived by the model fitter for all RMS sources with $\delta\bar{F}_{w}$~/~$\bar{F}_{w}$~$\leq$~0.5 and $\chi^{2}_{best}$~$\leq$~30 for $\Delta\chi^{2}$~$\leq$~10.}
\label{F:tests_extinction}
\end{figure}

We limit the $A_{V}$ to be within the range 0 to 50, in line with typical values for UC\HII{} regions \citep[e.g.][]{Hanson2002}, though as shown in Figure~\ref{F:tests_extinction} the majority of sources have fitted values below this.

\subsection{Data Included}
\label{S:tests_data}

\begin{figure}
\center
\includegraphics[width=0.45\textwidth]{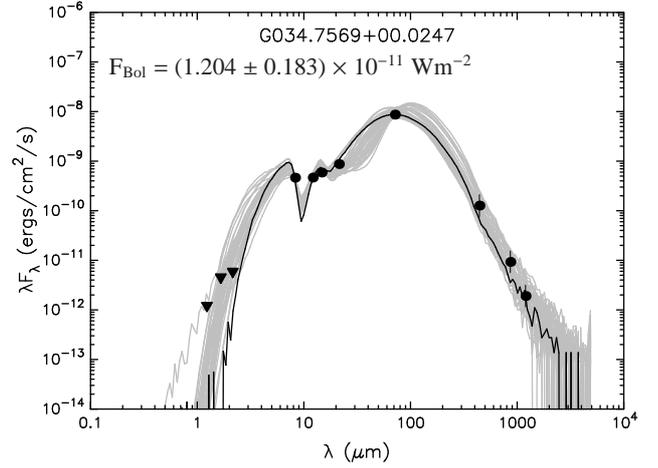}
\begin{picture}(290,1)(0,0)
\put(115,160){\makebox(0,0){F$_{\rm{Bol}}$~=~(1.204~$\pm$~0.183)~$\times$~10$^{-11}$~Wm$^{-2}$}}
\end{picture}
\includegraphics[width=0.45\textwidth]{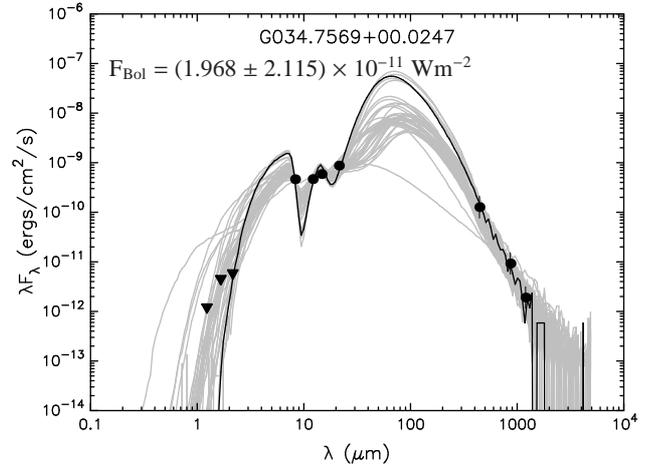}
\begin{picture}(290,1)(0,0)
\put(115,160){\makebox(0,0){F$_{\rm{Bol}}$~=~(1.968~$\pm$~2.115)~$\times$~10$^{-11}$~Wm$^{-2}$}}
\end{picture}
\includegraphics[width=0.45\textwidth]{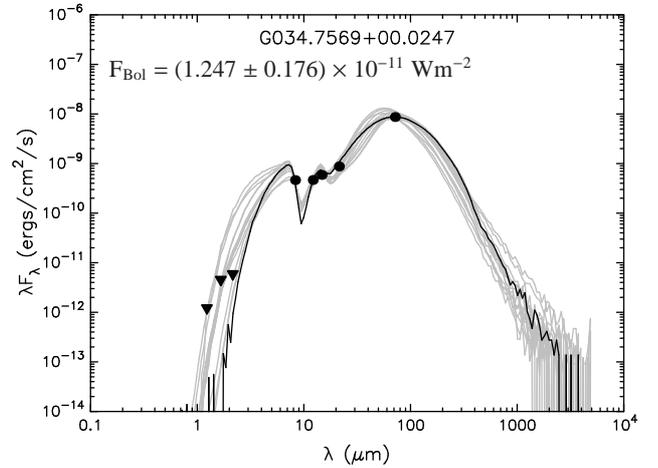}
\begin{picture}(290,1)(0,0)
\put(115,160){\makebox(0,0){F$_{\rm{Bol}}$~=~(1.247~$\pm$~0.176)~$\times$~10$^{-11}$~Wm$^{-2}$}}
\end{picture}
\caption{SED fits with different data sets included. Top: All data included. Middle: All data except MIPSGAL. Bottom: All data except sub-mm and mm. Bottom right: All data except 2MASS. The lines and points have the same meaning as in Figure~\ref{F:tests_distance_seds}. The bolometric flux derived for each data set is indicated on the plot.}
\label{F:tests_data_seds}
\end{figure}

SED model fits for G034.7569+00.0247 including only subsets of the available data were undertaken in order to explore the effect of missing data on the bolometric flux results, examples of which, along with the fit using all data are shown in Figure~\ref{F:tests_data_seds}. Comparing the top plot, which includes all data, with the middle plot where the MIPSGAL data has been omitted, the effect of using the model fitter for sources without far-IR data is revealed. The lack of constraint at the peak of the SED leads to the fractional error in the bolometric flux being larger than 1, i.e. it is unconstrained, despite having an otherwise reasonably complete data set. As a result, as mentioned in \S\ref{S:data} we do not fit SEDs for sources where we do not have far-IR flux data, though we will see in \S\ref{S:results_nofir} that we can use the F$_{21}$ to F$_{\rm{Bol}}$ ratio as derived from fitted sources to estimate F$_{\rm{Bol}}$ for these sources. The lower plot of Figure~\ref{F:tests_data_seds} illustrates that the lack of sub-mm and mm data for many of our sources does not change the results within the calculated errors.

\section{Results}
\label{S:results}

The SED model fitter of \citet[][]{Robitaille2007a} was run for all 1183 sources with far-IR data as discussed in \S\ref{S:data} using kinematic distances with $\delta$$d$~=~1~kpc obtained from molecular emission lines \citep[][Urquhart \etal{}, 2010b, in prep.]{Urquhart2007c,Urquhart2008a,Urquhart2010} toward each source. There are 33 sources remaining with multiple solutions for the kinematic distance to some sources due to unresolved near/far distance ambiguity, for which the SED fitter was run for each possible distance and the flux and error are given as the mean and $\sigma$/$\sqrt{n}$ of the results from these runs. The maximum $A_{V}$ allowed was 50 (see \S\ref{S:tests_extinction}).

\begin{table*}
\centering
\caption{Example results of SED fitting for all sources with far-IR data. The columns are as follows - 1: MSX PSC name. 2: Right ascension. 3:Declination. 4: RMS classification. 5: Number of detected flux data points, not including upper limits, included in the fit. 6: Flags indicating which data were included in the SED. These are from left to right 2MASS (Y or N), TIMMI2 (Y or N), far-IR IGA (I) or MIPSGAL (M), submillimetre (Y or N) and millimetre (N, 1=\citet{Beuther2002a}, 2=\citet{Faundez2004}, 3=\citet{Hill2005}, 4=\citet{Beltran2006}). 7: $\chi^{2}_{best}$ for the fit. 8: $\Delta\chi^{2}$ at which at least 10 fits are found. 9: The number of fits. 10: The bolometric flux with uncertainty, calculated using equations~\ref{E:tests_flux} and \ref{E:tests_weights}. 11: Whether the kinematic distance has been solved (s) or the near/far ambiguity remains unsolved (nf). Sources with galactocentric radii $\geq$~8.5~kpc or at the tangent point are also indicated as solved (s). 12: The apportioning correction applied to large-beam fluxes, as discussed in \S\ref{S:data_timmi2}, with a value of 1 meaning that no apportioning was performed. 13: Type of apportioning used with 'N' for none, 'T' for TIMMI2 data, 'G' for GLIMPSE data and 'S' for a simple split of the remaining fraction between all multiple source components without TIMMI2 or GLIMPSE detections. A full version of this Table is available online at the CDS via anonymous ftp to cdsarc.u-strasbg.fr (130.79.125.5) or via http://cdsweb.u-strasbg.fr/cgi-bin/qcat?J/A+A/.}
\begin{tabular}{@{}lcc@{~~}c@{~~}c@{~~}l@{~~}ccccccc@{}}
\hline
 Object & RA & Dec & RMS & $\#$ & \multicolumn{1}{c}{Data} & $\chi^{2}_{best}$ & $\Delta\chi^{2}$ & $\#$ & Bolometric Flux  & Type & Ap. & Ap. \\
 & (J2000) & (J2000) & Ident.  & Pts & \multicolumn{1}{c}{Incl.} & &  & Fits  & (10$^{-11}$ Wm$^{-2}$) & Cor. & Type \\
\hline
G010.3844+02.2128&18:00:22.67&$-$18:52:09.7&YSO& 8&YN~I~NN& 1.6& 2.0&  32&3.300$\pm$0.371&s& 1.00&N\\
G010.5067+02.2285&18:00:34.58&$-$18:45:17.6&YSO& 7&YN~I~NN& 2.0& 2.0& 183&0.618$\pm$0.146&s& 1.00&N\\
G010.6207$-$00.3199&18:10:14.11&$-$19:54:07.5&C\HII{}& 5&YNMNN& 4.8& 4.0&  11&1.350$\pm$0.130&s& 1.00&N\\
G010.6291$-$00.3385&18:10:19.25&$-$19:54:12.1&\HII{}& 6&YNMNN&14.2& 4.0&  11&3.901$\pm$0.319&s& 1.00&N\\
G010.8411$-$02.5919&18:19:12.09&$-$20:47:30.9&YSO&11&YN~I~Y3& 2.5& 2.0&  30&20.660$\pm$3.162&s& 1.00&N\\
G010.9657+00.0083&18:09:43.36&$-$19:26:28.6&\HII{}& 7&YNMNN& 6.1& 4.0&  10&1.497$\pm$0.155&s& 1.00&N\\
G011.1109$-$00.4001&18:11:32.28&$-$19:30:40.6&\HII{}& 6&YNMNN&11.1&12.0&  15&5.433$\pm$0.367&s& 1.00&N\\
G011.5001$-$01.4857&18:16:22.58&$-$19:41:19.3&\HII{}$/$YSO&10&YN~I~YN& 6.0& 2.0&  13&7.478$\pm$1.030&s& 1.00&N\\
G011.9019+00.7265&18:08:58.78&$-$18:16:29.8&Y$/$O Str& 7&YNMNN& 0.3& 2.0& 134&0.692$\pm$0.163&s& 0.88&G\\
G011.9454$-$00.0373&18:11:53.20&$-$18:36:21.8&\HII{}& 8&YNMN3& 2.0& 2.0&  33&6.329$\pm$2.001&s& 1.00&N\\
G013.0105$-$00.1267&18:14:22.17&$-$17:42:48.8&Y$/$O Str& 6&YNMNN& 2.1& 2.0&  44&3.106$\pm$1.220&s& 1.00&N\\
G013.1840$-$00.1069A&18:14:38.57&$-$17:33:05.8&YSO& 5&NNMNN& 4.0& 2.0&  15&0.308$\pm$0.048&s& 0.37&G\\
G013.1840$-$00.1069B&18:14:38.57&$-$17:33:05.8&\HII{}& 8&YNMNN& 8.6& 3.0&  15&0.420$\pm$0.061&s& 0.37&G\\
G013.2097$-$00.1436&18:14:49.94&$-$17:32:46.6&C\HII{}& 8&YNMNN&44.9&14.0&  10&5.084$\pm$0.504&s& 1.00&N\\
G015.0939+00.1913&18:17:20.87&$-$15:43:45.9&YSO& 6&YNMNN& 2.8& 3.0&  28&0.372$\pm$0.146&s& 0.65&G\\
G017.0332+00.7476A&18:19:07.33&$-$13:45:23.6&YSO& 5&NNMNN& 4.4& 2.0&  26&0.742$\pm$0.243&s& 0.29&G\\
G017.0332+00.7476B&18:19:07.22&$-$13:45:30.2&\HII{}& 8&YNMNN& 9.2& 2.0&  12&0.710$\pm$0.117&s& 0.71&S\\
\hline
\end{tabular}
\label{T:results_full}
\end{table*}

A sample of the results from these fits are shown in Table~\ref{T:results_full}, while the full version of this table is available online at the CDS via anonymous ftp to cdsarc.u-strasbg.fr (130.79.125.5) or via http://cdsweb.u-strasbg.fr/cgi-bin/qcat?J/A+A/. The RMS sample sources have been given one of a number of classifications, shown in column 4 of Table~\ref{T:results_full}, which are discussed in \S2.1 of \citet{Mottram2010}. `Young~/~Old Star' is abbreviated to `Y/O Str' and an additional classification for more evolved Compact or diffuse \HII{} regions, where the ionised region fills much of the MSX beam, is abbreviated to `C\HII{}'. For details on individual sources see the RMS database (http://www.ast.leeds.ac.uk/RMS). As distance solutions are subject to change as new information becomes available, these are not included in the table. However, up to date distance information is available on the database and once distances are available for all sources, these will be the subject of a future publication. In addition, as discussed in \S\ref{S:tests_distance} the flux results are distance independent. The data included in the SED of a source is given by a series of flags in column 6 of Table~\ref{T:results_full}. These are from left to right 2MASS (Y or N), TIMMI2 (Y or N), whether the far-IR was from the IGA (I) or MIPSGAL (M), submillimetre (Y or N) and millimetre data (N, 1=\citet{Beuther2002a}, 2=\citet{Faundez2004}, 3=\citet{Hill2005}, 4=\citet{Beltran2006}). Table~\ref{T:results_full} contains results for 1173 sources, while for a further 10 sources the fitter was run but returned a flux smaller than the error in the flux (i.e. unconstrained).

We do not discuss in this paper the results of the 14 free parameters in the models of \citet[][]{Robitaille2007a} as these are not uniquely constrained by the data, and the models do not cover the parameter space uniformly. Indeed, the distributions of most of the parameters follow trends in the models \citep[see][for more details]{Mottram2008}.

\begin{figure}
\center
\includegraphics[width=0.49\textwidth]{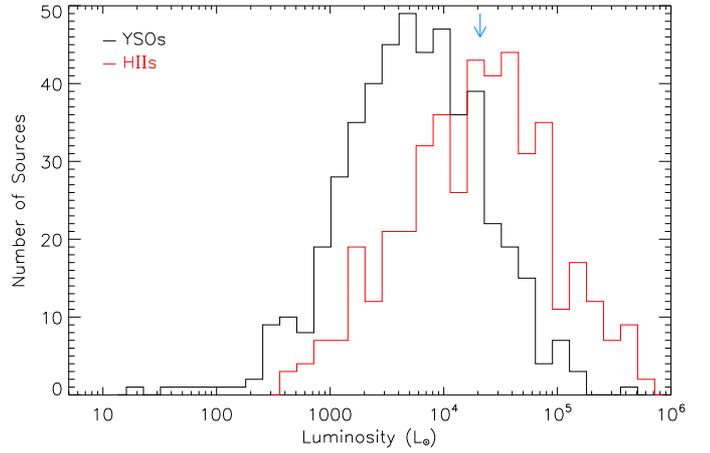}
\caption{Plots of the number of sources vs. luminosity for RMS candidates with classifications of YSO (488 sources, black) and \HII{} (440 sources, red). The blue arrows indicate the luminosity of a B0 type star from the calculations of \citet[][]{Martins2005} corrected for the luminosity/temperature/mass relationship of \citet[][]{Meynet2000}.}
\label{F:discussion_nplots}
\end{figure}

\begin{figure*}
\center
\includegraphics[width=0.49\textwidth]{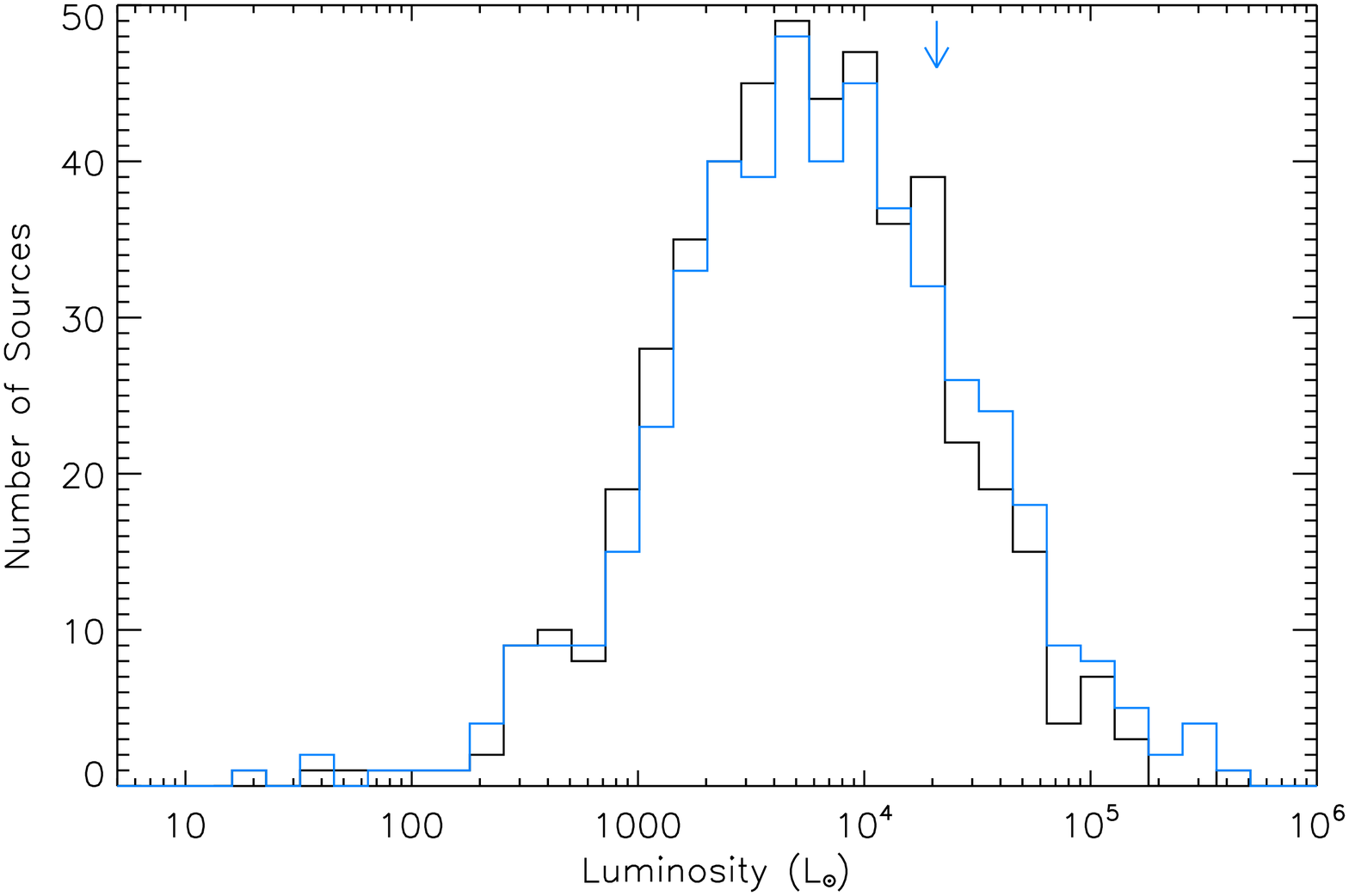}
\includegraphics[width=0.49\textwidth]{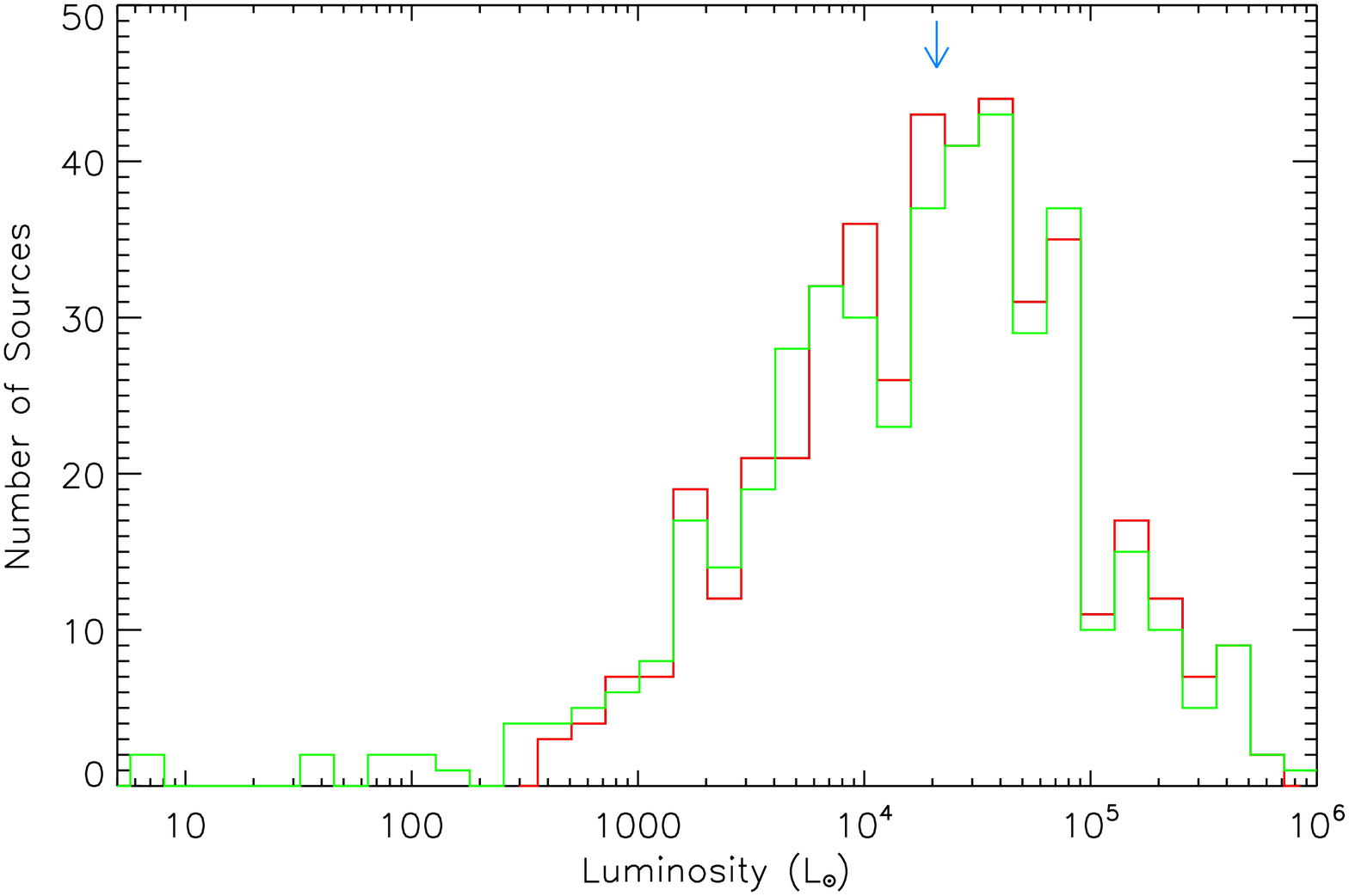}
\caption{Plots of the number of sources vs. luminosity for YSOs (left) and \HII{} regions (right) with the assigned distances (colours as in figure~\ref{F:discussion_nplots}) and with all near/far distance determinations swapped (blue and green respectively). The blue arrows have the same meaning as in Figure~\ref{F:discussion_nplots}}
\label{F:discussion_nfswitch}
\end{figure*}

A total of 1069 sources have an unique assigned kinematic distance and have fits with $\delta$$\bar{F_{w}}$~$/$~$\bar{F_{w}}$~$\leq$~0.5 and $\chi^{2}_{best}$~$\leq$~30 for $\Delta\chi^{2}$~$\leq$~10, of which 488 are identified as YSOs, 20 as \HII{}$/$YSOs, 45 as Young~/~Old Star, 76 as C\HII{} and 440 as \HII{} regions. Histograms of the number of sources as a function of their luminosity for YSOs and \HII{} regions are shown in Figure~\ref{F:discussion_nplots}. These sources have typical uncertainties in the bolometric flux of $\sim$~17$\%$, have a median distance of 5.1~kpc, a typical distance uncertainty of $\sim$~19~$\%$ (assuming $\delta$d~=~1.0~kpc), and have typical uncertainties in the luminosity of 34$\%$. The sources with MIPSGAL far-IR fluxes do not show any significant difference in terms of distribution to those with IGA data, implying that the improvement in resolution provided by the MIPSGAL data does not result in a lack of luminous sources. Figure~\ref{F:discussion_nfswitch} shows the plots for YSOs and \HII{} regions from Figure~\ref{F:discussion_nplots} for both the assigned distances (black) and with all near/far determinations for all assigned kinematic distances inverted. This is undertaken to examine the effect of incorrect near/far distance ambiguity assignments on the luminosity distributions. As can be see, this does not change the overall shape of the luminosity distributions appreciably. Therefore the remaining individual uncertainties in the distances to the RMS sample do not significantly affect the global results on a statistical level. These results will be discussed in detail in \S\ref{S:discussion}.

\subsection{Sources Without Far-IR Data}
\label{S:results_nofir}

\begin{figure*}
\center
\includegraphics[width=0.49\textwidth]{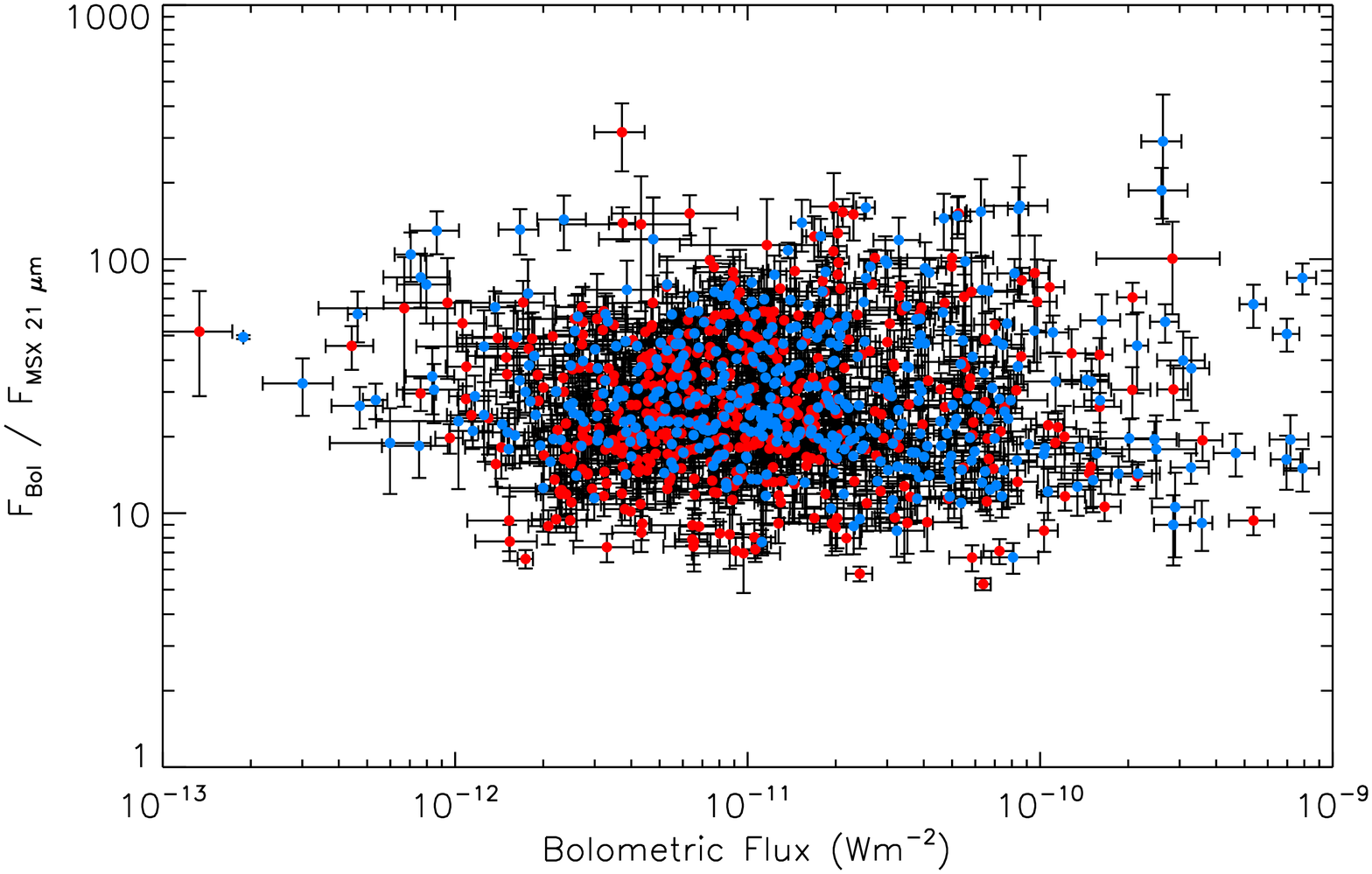}
\includegraphics[width=0.49\textwidth]{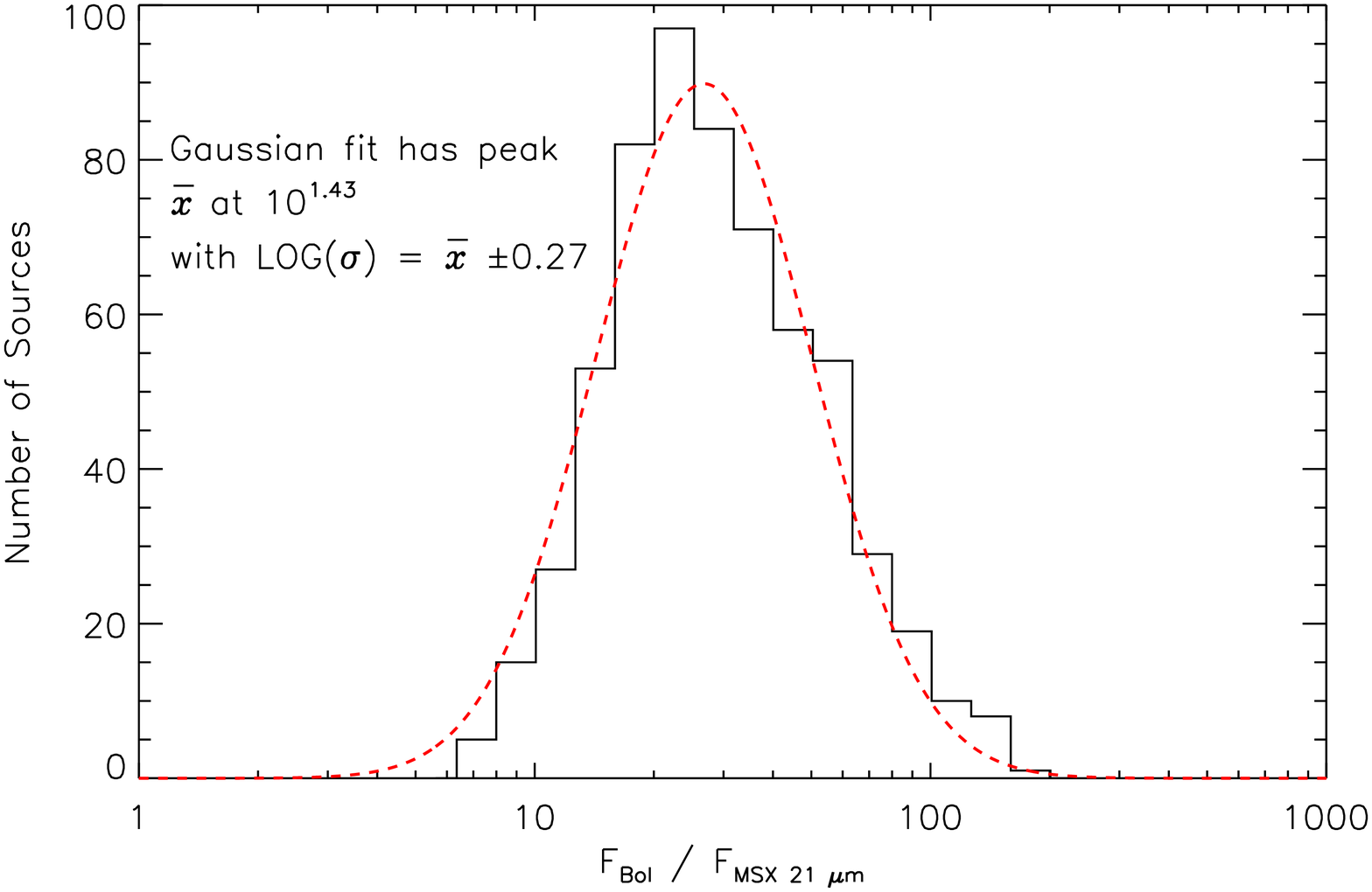}
\caption{Left: The distribution of $F_{\rm{Bol}}$~$/$~$F_{MSX~21}$ as a function of $F_{\rm{Bol}}$ for all YSOs and \HII{} regions (red and blue respectively in the online version of this Figure). Error bars are shown in black in order to give an idea of the total variation in values. Right: The histogram of $F_{\rm{Bol}}$~$/$~$F_{MSX~21}$ for all sources with MIPS data. A Gaussian is fitted to the data is shown by the dashed line.}
\label{F:results_nofir_ftotf21}
\end{figure*}

Though far-IR fluxes were obtained by \citet[][]{Mottram2010} for the majority of young RMS sources, this was not possible for some sources due to confusion within the available data, which in turn prevented SED fitting of these sources (see \S\ref{S:tests_data}). However, the properties of those sources where fits were obtained can be used to calculate the mean ratio between the bolometric flux and the MSX 21\micron{} flux for all sources with good SED fits. This mean ratio can then be used to obtain estimates of the bolometric flux of RMS sources which do not have far-IR data, since all have measured MSX 21\micron{} fluxes.

In addition to the total flux, filter-band fluxes were output for the SED model fits to the data. The MSX 21\micron{} band filter fluxes in Wm$^{-2}$ can be calculated by multiplying the SED MSX 21\micron{} flux in Janskys by the bandwidth of the filter \citep[4.041~$\times$~10$^{-14}$~Hz,][]{Cohen2000}. The weighted mean ratio of the total bolometric flux to the MSX 21\micron{} band filter flux can therefore be obtained for each source, in a similar manner to that used for the bolometric flux (see \S\ref{S:tests}), with the uncertainty given by the weighted standard deviation. The results of these calculations do not appear to be dependent on source flux or type (see left-hand plot of Figure~\ref{F:results_nofir_ftotf21}), though the scatter in sources is relatively large and so may mask such relationships. The ratio itself has a roughly log-normal distribution, as shown in the right-hand plot of Figure~\ref{F:results_nofir_ftotf21}. Sources with MIPSGAL data generally have a slightly smaller spread than the general distribution, so the properties of the Gaussian fits to the mean ratio for these 613 sources are used.

This mean ratio was therefore used to obtain estimates of the bolometric flux of RMS sources which do not have far-IR data, since all have measured MSX 21\micron{} fluxes. The bolometric fluxes obtained for sources with TIMMI2 (69) or GLIMPSE PSC (20) fluxes, as well as split sources (19), were apportioned using the method discussed in \S\ref{S:data_timmi2}. An example of these results is presented in Table~\ref{T:nofir_results_full}, while the full version of this table is available online at the CDS via anonymous ftp to cdsarc.u-strasbg.fr (130.79.125.5) or via http://cdsweb.u-strasbg.fr/cgi-bin/qcat?J/A+A/.

Estimate total fluxes were obtained using this method for an additional 280 young RMS sources. The uncertainty in the total flux for each source is a combination in quadrature of the uncertainty in $F_{\rm{Bol}}$~$/$~$F_{MSX~21}$ and the uncertainty in the MSX 21\micron{} flux. The fact that the Gaussian fit to $F_{\rm{Bol}}$~$/$~$F_{MSX~21}$ is performed in log space results in asymmetric uncertainties, which are propagated through to the total flux.

\begin{table*}
\centering
\caption{Example bolometric fluxes for young RMS sources without far-IR data. A full version is available online at the CDS via anonymous ftp to cdsarc.u-strasbg.fr (130.79.125.5) or via http://cdsweb.u-strasbg.fr/cgi-bin/qcat?J/A+A/.}
\begin{tabular}{@{~}lccccccc@{~}}
\hline
 Object & RA & Dec & RMS & MSX 21~\micron{} & Bolometric Flux & Ap. & Ap. \\
 & (J2000) & (J2000) & Ident.  & Flux (Jy)  & (10$^{-11}$ Wm$^{-2}$) & Cor. & Type \\
\hline
G010.1615$-$00.3623&18:09:26.88&$-$20:19:28.2&C\HII{}& 692.20$\pm$  41.53&75.287$\pm$$^{65.061}_{35.15}$&1.00&N\\
G010.1880$-$00.3162&18:09:19.82&$-$20:16:44.4&C\HII{}&   9.63$\pm$   0.59&1.048$\pm$$^{0.905}_{0.49}$&1.00&N\\
G010.4413+00.0101&18:08:38.23&$-$19:53:57.4&\HII{}&   9.91$\pm$   0.59&1.078$\pm$$^{0.932}_{0.50}$&1.00&N\\
G010.4718+00.0206&18:08:39.67&$-$19:52:03.0&\HII{}&   3.84$\pm$   0.24&0.418$\pm$$^{0.361}_{0.20}$&1.00&N\\
G010.6311$-$00.3864&18:10:30.26&$-$19:55:30.0&\HII{}&  15.25$\pm$   0.92&1.659$\pm$$^{1.433}_{0.77}$&1.00&N\\
G011.3757$-$01.6770&18:16:50.47&$-$19:53:20.4&\HII{}&   6.22$\pm$   0.39&0.676$\pm$$^{0.584}_{0.32}$&1.00&N\\
G011.4201$-$01.6815&18:16:56.85&$-$19:51:07.2&\HII{}$/$YSO&  98.36$\pm$   5.90&10.698$\pm$$^{9.245}_{4.99}$&1.00&N\\
G012.7408+00.3305&18:12:08.56&$-$17:43:50.8&C\HII{}&   3.40$\pm$   0.22&0.370$\pm$$^{0.320}_{0.17}$&1.00&N\\
G012.7890$-$00.2185&18:14:15.88&$-$17:57:06.1&C\HII{}&  20.27$\pm$   1.22&2.205$\pm$$^{1.905}_{1.03}$&1.00&N\\
G012.8062$-$00.1987&18:14:13.56&$-$17:55:37.5&\HII{}& 679.40$\pm$  40.76&73.895$\pm$$^{63.858}_{34.50}$&1.00&N\\
G012.8909+00.4938A&18:11:51.13&$-$17:31:21.1&YSO&  12.93$\pm$   0.78&0.585$\pm$$^{2.034}_{0.18}$&0.42&G\\
G012.8909+00.4938B&18:11:50.68&$-$17:31:14.5&\HII{}&  12.93$\pm$   0.78&0.680$\pm$$^{1.940}_{0.11}$&0.48&S\\
G012.8909+00.4938C&18:11:51.44&$-$17:31:30.1&YSO&  12.93$\pm$   0.78&0.142$\pm$$^{2.477}_{0.61}$&0.10&G\\
G014.9790$-$00.6649A&18:20:15.60&$-$16:14:10.3&YSO&  54.24$\pm$   3.25&2.950$\pm$$^{8.043}_{0.41}$&0.50&S\\
G014.9790$-$00.6649B&18:20:15.60&$-$16:14:10.3&\HII{}&  54.24$\pm$   3.25&2.950$\pm$$^{8.043}_{0.41}$&0.50&S\\
G014.9958$-$00.6732&18:20:19.47&$-$16:13:29.8&YSO&  98.85$\pm$   5.93&10.751$\pm$$^{9.291}_{5.02}$&1.00&N\\
G015.0136$-$00.7037&18:20:28.24&$-$16:13:26.4&C\HII{}& 376.60$\pm$  22.60&40.961$\pm$$^{35.397}_{19.12}$&1.00&N\\
\hline
\end{tabular}
\label{T:nofir_results_full}
\end{table*}

\begin{figure*}
\center
\includegraphics[width=0.49\textwidth]{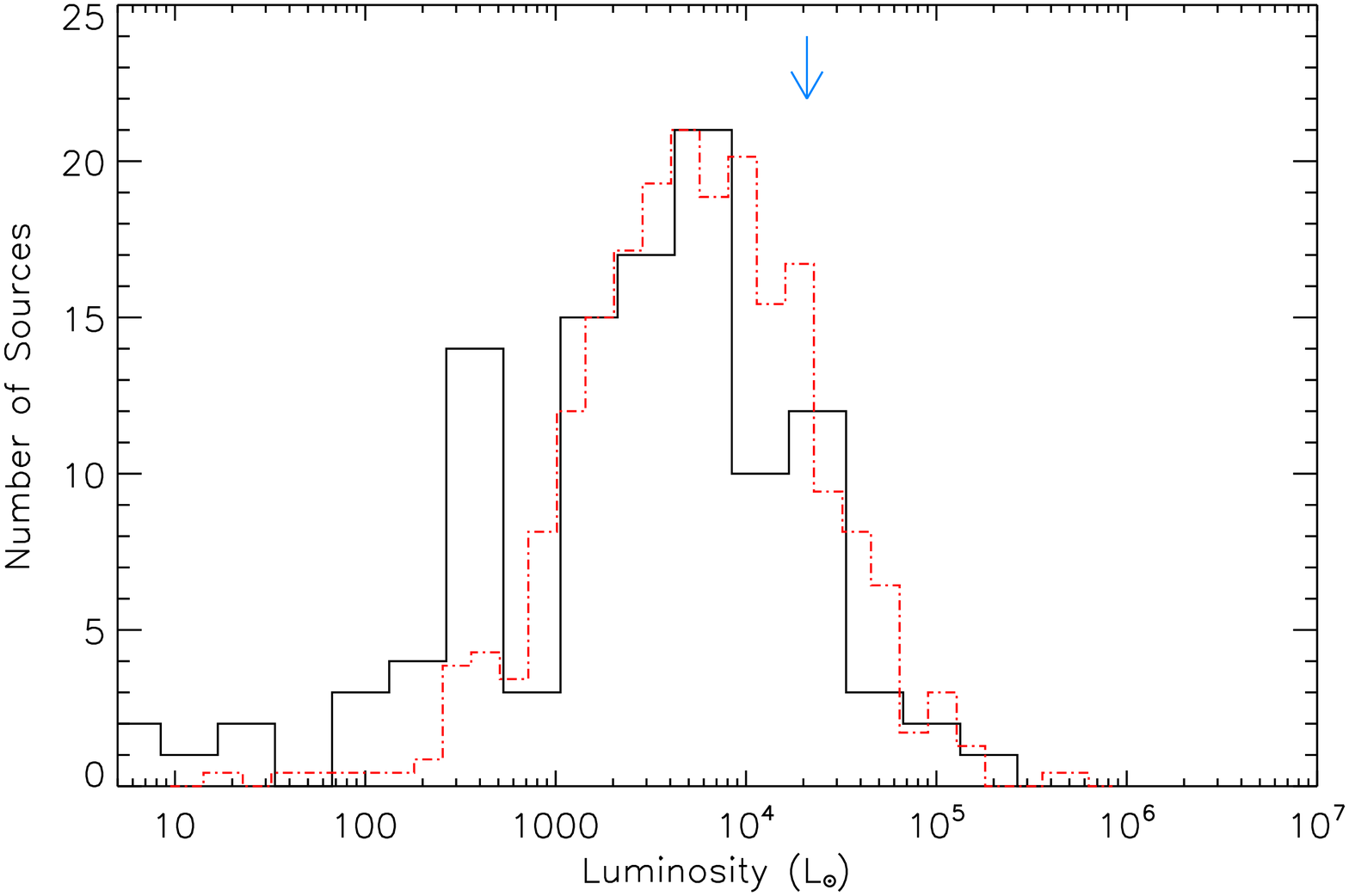}
\includegraphics[width=0.49\textwidth]{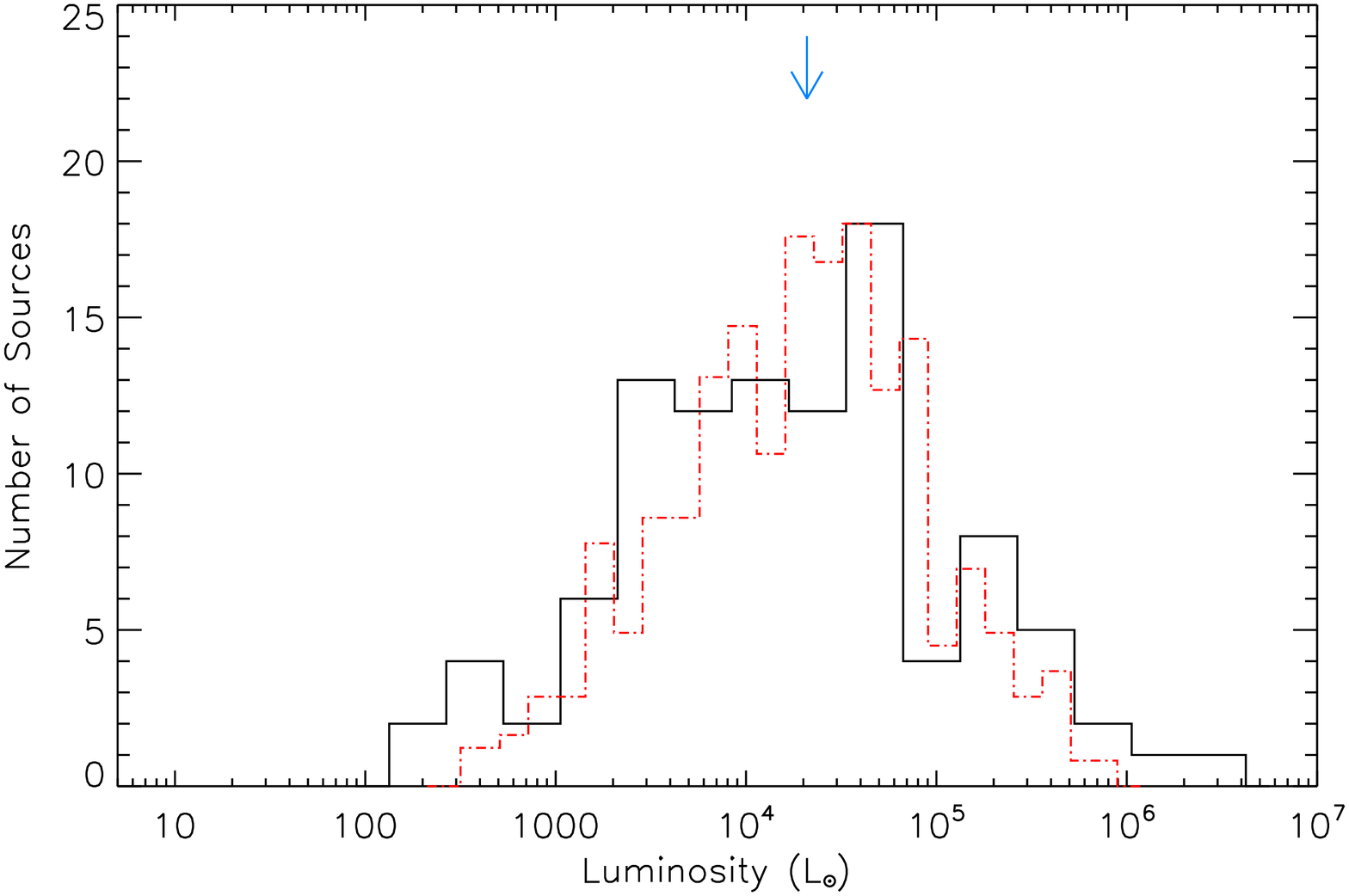}
\caption{Plots of the number of sources vs. luminosity calculated from MSX 21\micron{} fluxes for RMS candidates without far-IR data with classifications of YSO (110 sources, left) and  \HII{} region (103 sources, right). The blue arrows have the same meaning as in Figure~\ref{F:discussion_nplots}. The dot-dashed histograms show the normalised luminosity distributions from Figure~\ref{F:discussion_nplots}.}
\label{F:results_nofir_nplots}
\end{figure*}

The distributions of luminosities for sources without far-IR data identified as YSO or \HII{} region with uniquely assigned distances is shown in Figure~\ref{F:results_nofir_nplots}, with the distributions of SED derived luminosities from Figure~\ref{F:discussion_nplots} also shown for comparison. Due to the lower number of sources (110 YSOs and 103 \HII{} regions), the bin size for the distributions for luminosities derived using $F_{\rm{Bol}}$~$/$~$F_{MSX~21}$ is twice that used for the SED luminosities. The spread of sources is larger than for the SED luminosities, probably caused by the larger uncertainties in the bolometric fluxes derived using the MSX 21\micron{} flux and the fact that these sources are in regions which are confused at far-IR wavelengths. However the general shape of the distributions is similar, and the peaks in the distributions are near those for the SED luminosities. This method of estimating the far-IR flux therefore seems to provide a reasonable estimate when no far-IR information is available.
 
\subsection{Filter Effects}
\label{S:results_filters}

As with the ratio of $F_{\rm{Bol}}$~$/$~$F_{MSX~21}$ discussed above, various other ratios both of filter fluxes to the bolometric flux and filter fluxes to other filter fluxes can be calculated from the model SED fits. However the SED models of \citet[][]{Robitaille2006} do not include PAH emission features, so the flux ratios for bands where these are important may be different, e.g. the 8.0\micron{} PAH band may affect ratios involving the MSX~8\micron{} band. The results of these fits are presented in Table~\ref{T:results_filters}.

\begin{table}
\centering
\begin{tabular}{@{~}ccccc@{~}}
\hline
\centering
Flux Ratio & \={x} & $\sigma$ & \\
\hline
Log$_{10}(F_{\rm{Bol}}$~$/$~$F_{MSX~21})$ & 1.43 & 0.27 \\
Log$_{10}(F_{\rm{Bol}}$~$/$~$F_{TIMMI2~10.4})$ & 2.53 & 0.38 \\
Log$_{10}(F_{MSX~21}$~$/$~$F_{MSX~8})$ & 0.89 & 0.32 \\
Log$_{10}(F_{MIPS~70}$~$/$~$F_{MSX~21})$ & 1.14 &  0.35 \\
Log$_{10}(F_{IRAS~60}$~$/$~$F_{MIPS~70})$ & -0.04 &  0.07 \\
Log$_{10}(F_{IRAS~100}$~$/$~$F_{MIPS~70})$ & 0.05 & 0.08 \\
\hline
\end{tabular}
\caption{Filter flux ratios derived from the SED fits to sources with MIPSGAL detections.}
\label{T:results_filters}
\end{table} 

\section{Discussion}
\label{S:discussion}

\subsection{Comparison with Other Methods}
\label{S:discussion_other}

\begin{figure*}
\center
\includegraphics[width=0.49\textwidth]{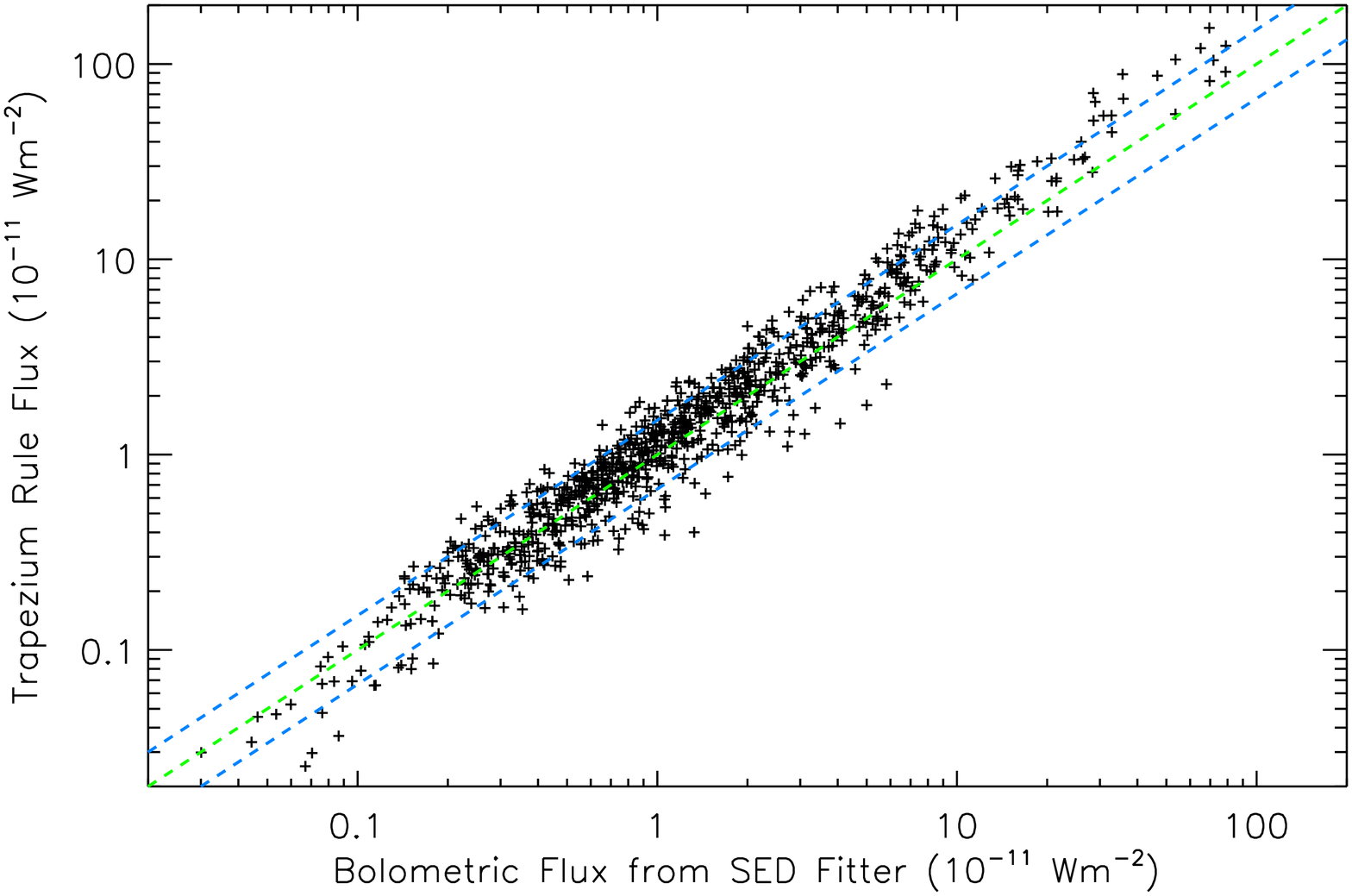}
\includegraphics[width=0.49\textwidth]{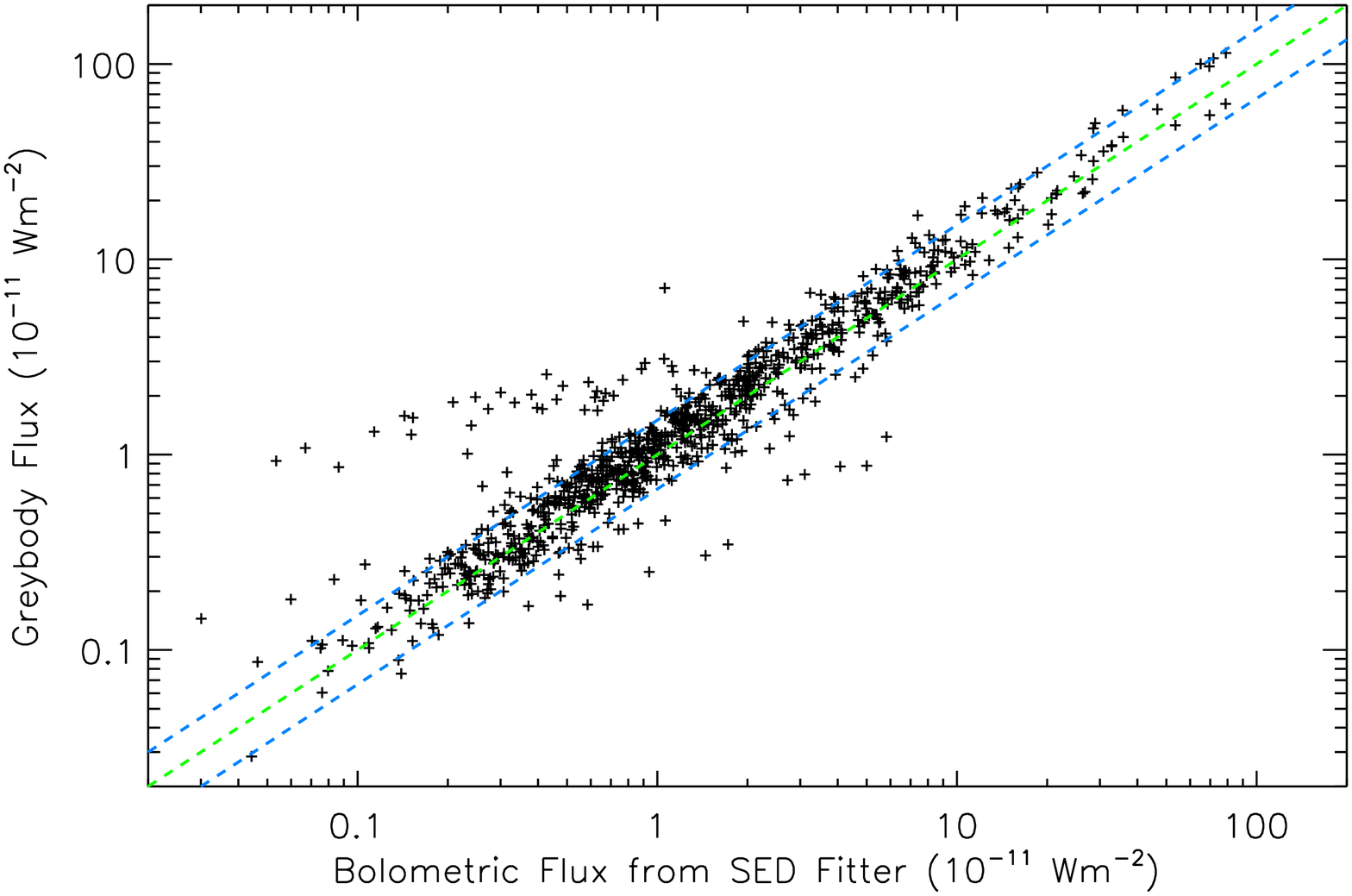}
\caption{Comparison of bolometric fluxes derived by the model SED fitter with those obtained by simple trapezium rule integration (left) and two component greybody fits (right). The green dashed line indicates y~=~x. The blue dashed lines indicate equality $\pm$~50$\%$.}
\label{F:discussion_other_seds}
\end{figure*}

\begin{figure*}
\center
\includegraphics[width=0.46\textwidth]{G013.6562-00.5997_sed.eps}
\includegraphics[width=0.49\textwidth]{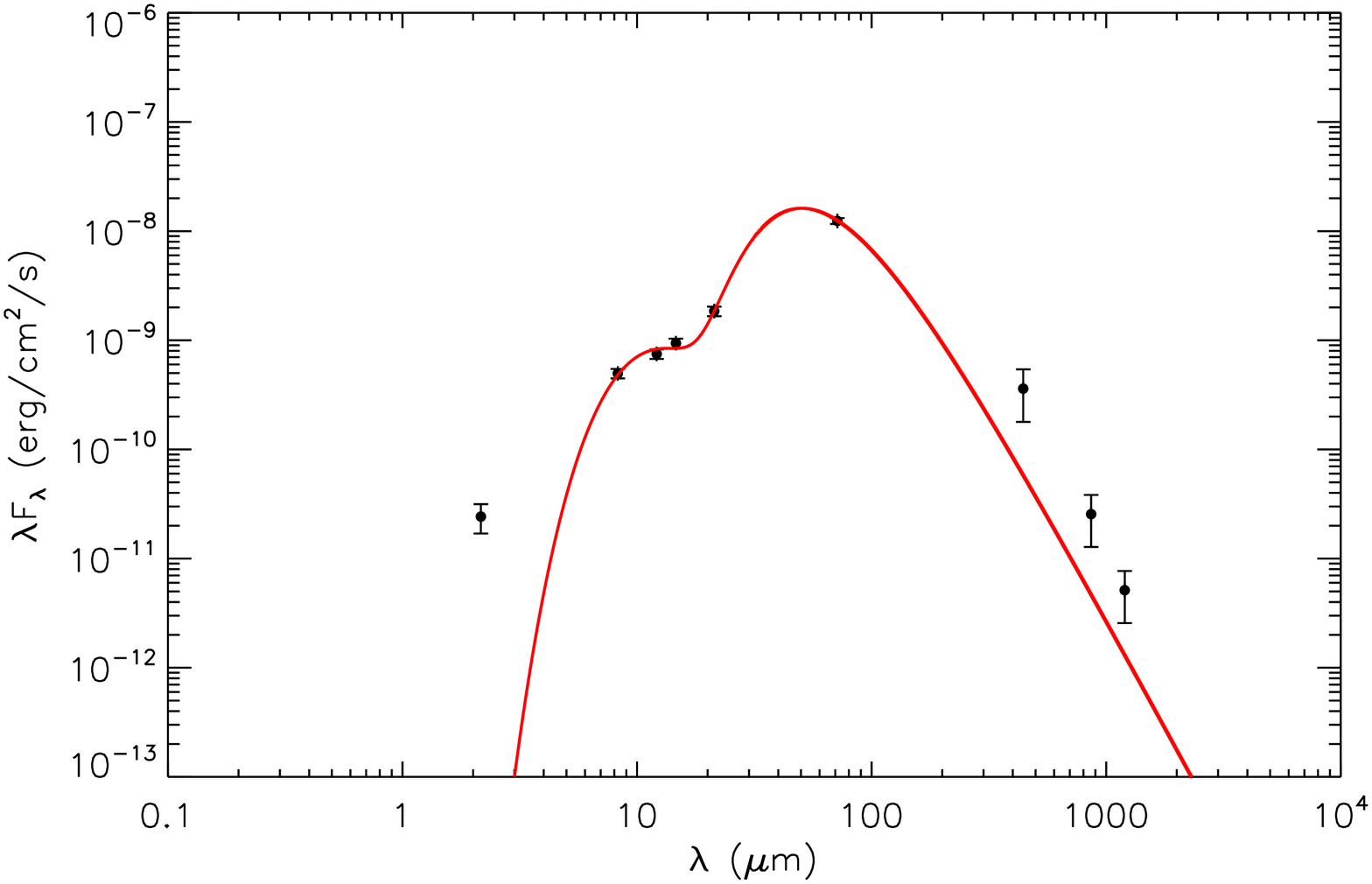}
\caption{Example SED (left) and greybody (right) fits to the same data. For the SEDs, the lines and points have the same meaning as in Figure~\ref{F:tests_distance_seds}. The solid line in the greybody images indicates the best two-component greybody fit to the data.}
\label{F:discussion_other_greybodyseds}
\end{figure*}

In order to explore the relative accuracy of other methods commonly used to obtain bolometric fluxes compared to the SED fitter results, we performed calculations using both simple trapezium rule integration and a combination of two greybody fits \citep[e.g.][]{Minier2005,Hill2009} with model rather than observed fluxes. The SED model `observed' flux for each fit was first corrected for the extinction derived by that fit. Next the weighted mean and weighted standard deviation were calculated in order to obtain a corrected model flux with error for each filter where data were input to the fitter for that source. The greybody fit was obtained using two temperature components given by:

\begin{equation}
\begin{array}{c}
F_{\nu}~=~B_{\nu}(T_{1})(1~-~e^{-\tau_{\nu1}})~\theta_{1}^{2}+B_{\nu}(T_{2})(1~-~e^{-\tau_{\nu2}})~\theta_{2}^{2}\vspace{1mm}\\ 
\mbox{where}~~B_{\nu}(T)~=~\frac{2h\nu^{3}}{c^{2}\left(e^{\frac{h\nu}{kT}}-1\right)}~~\mbox{and}~~\tau_{\nu}~=~\tau_{ref}(\nu~/~\nu_{ref})^{\beta}
\end{array}\vspace{1mm}
\label{E:discussion_other_greybody}
\end{equation}

\noindent
where $\theta$ is the source size, T is the dust temperature, $\tau_{\nu}$ is the optical depth at frequency $\nu$ and $\beta$ is the dust emissivity index such that $\tau$ can be evaluated at any frequency relative to a reference frequency $\nu_{ref}$. A total of five free parameters were used for the fits: the temperatures corresponding to the peaks of the two greybody components at lower and higher temperature (T$_{1}$ and T$_{2}$), the reference optical depth  ($\tau_{ref}$) at 850\micron{}, the source size in arcseconds of the high-temperature component ($\theta_{2}$) and $\beta_{1}$. $\beta_{2}$ was kept constant at 1, while $\theta_{1}$, the source size of the low-temperature component, was set to the beam size of 1.2~mm SEST data (i.e. 24\arcsec{}). $\beta_{1}$ was restricted to only be between 1 and 2. Only sources with at least five data points were fitted using this method, due to the number of free parameters, but this resulted in the exclusion of only a few sources which tend to have only upper limit 2MASS fluxes, an upper limit in the MSX 12\micron{} band and MIPSGAL data. Though these sources have less than 5 detected data points, the SED fitter is also constrained by the upper limits.

\begin{figure}
\center
\includegraphics[width=0.49\textwidth]{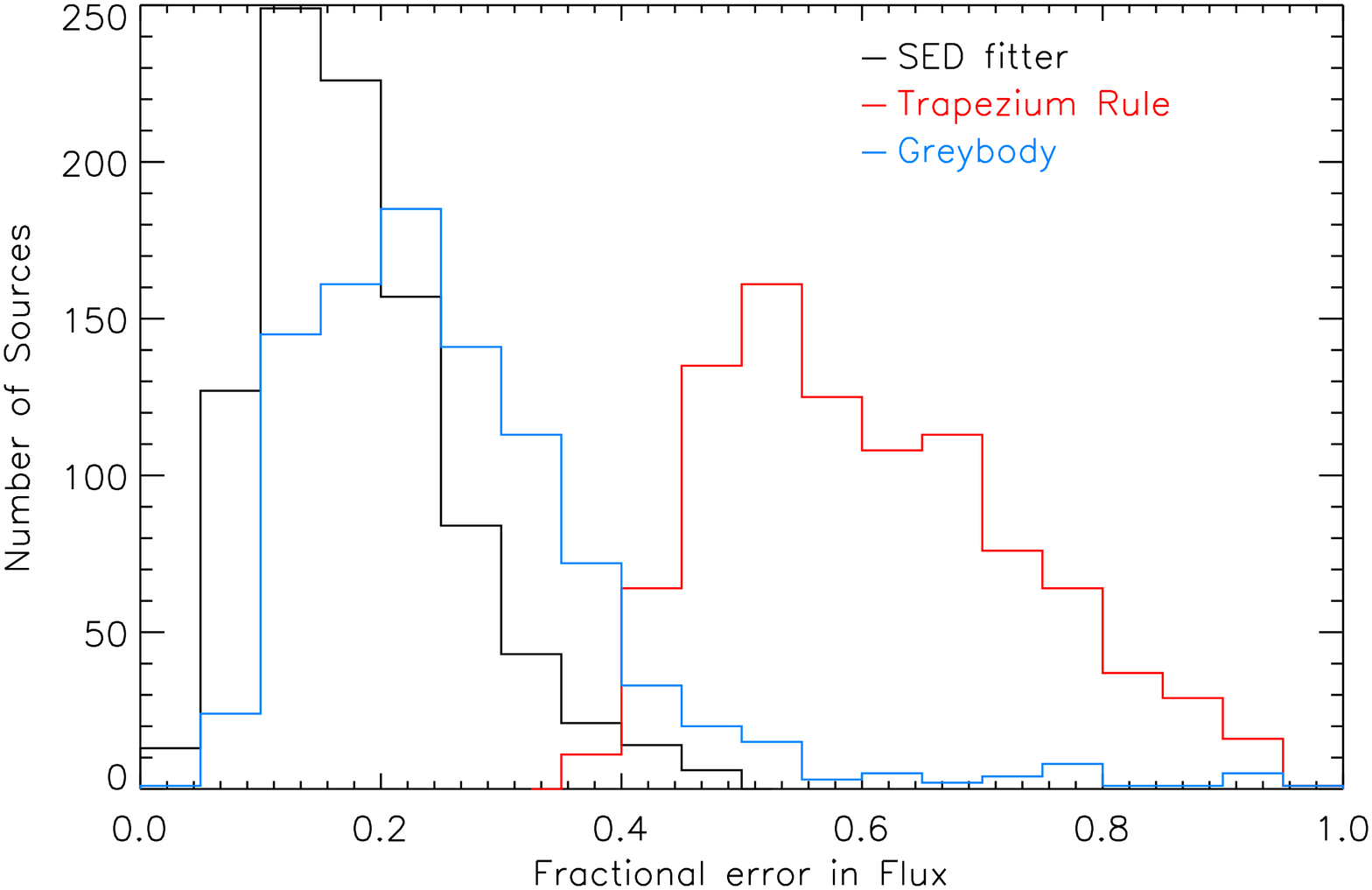}
\caption{Histogram of the fractional errors (i.e. $\delta$F$_{\rm{Bol}}$~/~F$_{\rm{Bol}}$) for fluxes obtained using the SED fitter (black), trapezium integration (red) and greybody fitting (blue).}
\label{F:discussion_other_errors}
\end{figure}

Figure~\ref{F:discussion_other_seds} shows a comparison between bolometric fluxes derived by the model SED fitter with those obtained by simple trapezium rule integration (left) and two component greybody fits (right). There is quite a large scatter, with a significant fraction more than 50$\%$ away from the SED fitter derived flux. Both alternative methods are more likely to over-estimate the flux, with mean values of F$_{grey}$~/~F$_{fitter}$~=~1.35 and F$_{trap}$~/~F$_{fitter}$~=~1.16. If the Emerson flux \citep{Emerson1988} is calculated with IRAS fluxes from the models instead of a full trapezium integration, the mean value of F$_{Emerson}$~/~F$_{fitter}$~=~1.09 though this does not include the effect of contamination which is often a major factor for IRAS fluxes of sources in the galactic plane. If IRAS PSC fluxes are used,  F$_{Emerson}$~/~F$_{fitter}$~=~1.82 with considerably more scatter. Many of the greybody fits do not fit both the near and mid-IR well, as shown by the example in Figure~\ref{F:discussion_other_greybodyseds}. The group of sources with considerably larger greybody fluxes in Figure~\ref{F:discussion_other_greybodyseds} are all sources with MIPS far-IR data but no submm or mm points where the fit has the far-IR peak at $\lambda$~$>$~70\micron{}, and so overestimates the flux from cold dust. Neither method will deal with the 9.7\micron{} silicate feature accurately. Figure~\ref{F:discussion_other_errors} shows a histogram of the fractional errors of fluxes derived by the SED fitter, greybody fits and trapezium integration. These indicate that errors in fluxes derived using trapezium integration are systematically larger than those from obtained using the other methods, and that the errors using greybody fits are on average larger than those using the SED fitter. The median fractional error is 0.17 for the SED fitter fluxes, 0.24 for the two-component greybody fluxes and 0.59 for trapezium integration. Overall, therefore, the SED fitter produces more consistent and accurate results than either greybody fitting of simple integration.

\subsection{YSOs}
\label{S:discussion_ysos}

The upper left plot of Figure~\ref{F:discussion_nplots} shows that the number of YSO sources peaks around 5~$\times$~10$^{3}$~\lsol{}, and decreases sharply above $\sim$~2~$\times$~10$^{4}$~\lsol{}. The gradual drop in the number of YSOs at luminosities below the peak is caused by incompleteness in the RMS sample due to the 21\micron{} flux-limit of the MSX survey. Though there are some sources which have L~$\gtrsim$~10$^{5}$~\lsol{} (which corresponds to an $\sim$~30~\msol{} O7 type star using the temperature/logQ scale of \citet{Martins2005} corrected for the luminosity/temperature/mass relationship of \citet{Meynet2000}), inspection reveals that many of these sources should be apportioned based on GLIMPSE images, and thus have lower luminosities, but lack PSC detections at 8\micron{}. As already discussed in \S\ref{S:results}, the shape of the overall distribution is robust against these issues.

\subsection{\HII{} Regions}
\label{S:results_hiis}

The left-hand plot of Figure~\ref{F:discussion_nplots} shows that the number of \HII{} regions in the RMS sample peaks around 3~$\times$~10$^{4}$~\lsol{}. There are few sources which are identified as \HII{} regions but have a luminosity derived from the SED lower than the limit for formation of an \HII{} region \citep[ L~$\sim$~550~\lsol{}, spectral type $\sim$B3,][]{Martins2005,Meynet2000}. Inspection of these sources suggest that they are correctly identified as \HII{} regions, but suffer from individual distance determination issues.

\begin{figure*}
\center
\includegraphics[width=0.49\textwidth]{G045.4543+00.0600.eps}
\includegraphics[width=0.49\textwidth]{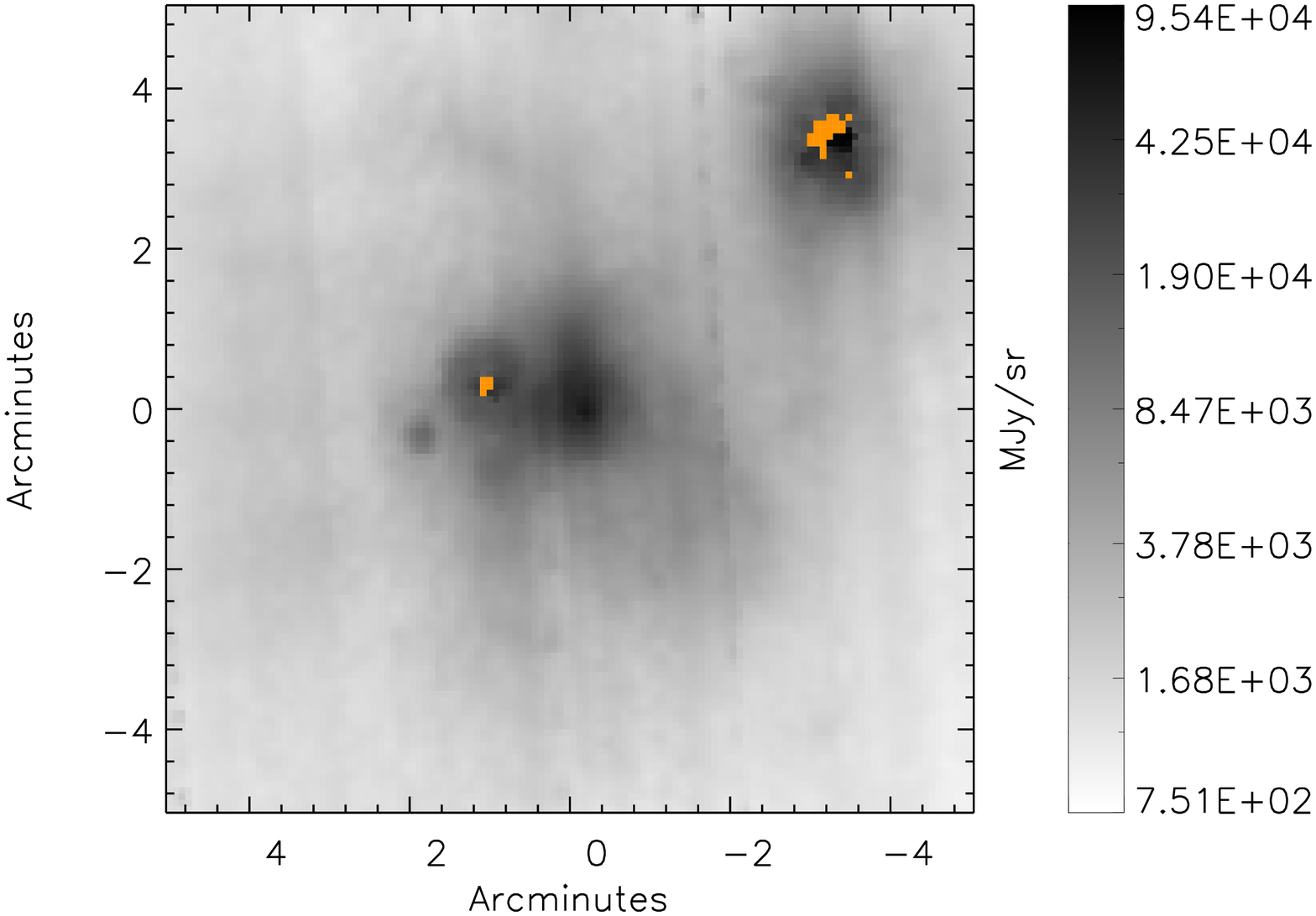}
\caption{SED (left) and MIPSGAL 70\micron{} image (right) for the \HII{} region G045.4543+00.0600, for which we derive a luminosity of (3.7~$\pm$~1.0)~$\times$~10$^{5}$~\lsol{} while \citet{Wood1989a} derive a luminosity, after correction to our distance of 7.3~kpc, of 8.2~$\times$~10$^{5}$~\lsol{} using IRAS far-IR data. The RMS source is merged with the source to the east (IRAS 19120+1103) in IRAS data and has a PSC 60\micron{} flux of 5340~Jy compared to a MIPS 70\micron{} flux of 2402~Jy, accounting for much of the remaining difference in luminosity measurements. Saturated pixels are indicated in yellow in the online version of this Figure. For the SED, the lines and points have the same meaning as in Figure~\ref{F:tests_distance_seds}.}
\label{F:discussion_hiis}
\end{figure*}

While the models used by the SED fitter do not include additional sources of heating of the dust inside the ionised zone, in particular Lyman alpha \citep[e.g.][]{Hoare1991}, the SEDs of UC\HII{} regions are similar to those of YSOs except at radio wavelengths, which the models do not probe. As the primary concern of this work is the bolometric flux, the fits are acceptable for this purpose.

The most luminous UC\HII{} region observed by \citet{Wood1989a} has a luminosity of $\sim$3~$\times$~10$^{6}$~\lsol{}, while no RMS source has a luminosities above 10$^{6}$~\lsol{}. However, some of the \citet{Wood1989a} sources are now known to contain multiple objects and in some cases the RMS kinematic distance measurements are significantly less than the ones used by those authors. For example \citet{Wood1989a} use d~=~9.7~kpc for G045.4543+00.0600 but the RMS kinematic distance to this source is 7.3~kpc, which would change their 1.44~$\times$~10$^{6}$~\lsol{} to 8.2~$\times$~10$^{5}$~\lsol{}. The SED fit for this source gives L~=~(3.7~$\pm$~1.0)~$\times$~10$^{5}$~\lsol{} but MIPSGAL rather than IRAS far-IR fluxes are used and the MIPSGAL 70\micron{} image shows multiple bright regions nearby (see Figure~\ref{F:discussion_hiis}). The SED does not fit the 850\micron{} point well, but this source exhibits significant surrounding low-level emission in the MIPSGAL image and strong radio emission. Both of these factors could increase the observed SCUBA flux above that consistent with the SED models based on the lower wavelength emission, which is well fit.

\subsection{Comparison of YSOs and \HII{} Regions}
\label{S:discussion_ysovshii}

In comparing the luminosity distributions of YSOs and \HII{} regions (Figure~\ref{F:discussion_nplots}), the difference in the peak of these distributions is apparent. Indeed, a Kolmogorov-Smirnov (KS) test of the cumulative distribution functions, indicates that the probability of these distributions being the same is of the order 10$^{-26}$. It is unlikely that this is due to an under-detection of B spectral type \HII{} regions in the sample as the RMS radio observations \citep{Urquhart2007a,Urquhart2009} were deep enough to detect \HII{} regions down to spectral type B1 to distances of 20~kpc assuming an electron temperature of 10$^{4}$~K. In addition, many of these MYSOs are visible at near-IR wavelengths, so are unlikely to be heavily embedded hyper-compact (HC) \HII{} regions.

Main-sequence stars with a spectral type of B3 (L~$\geq$~550\lsol{}) or earlier are capable of producing \HII{} regions, and evolutionary tracks indicate that core hydrogen burning should have started in YSOs of spectral type B3 or earlier despite the fact that accretion is still ongoing \citep[e.g.][]{Yorke2002}. It has therefore been unclear until recently why MYSOs, where the central source is massive enough to form an \HII{} region but no \HII{} region is observed, are detected at all, rather than all young massive stars transitioning directly from intermediate mass YSOs to UC\HII{} regions as they gain mass \citep[][]{Hoare2007b}. Some mechanism must either quench or mask the existence of an \HII{} region \citep[e.g.][]{Walmsley1995,Keto2003}, or increase the radius of the star such that the stellar T$_{eff}$, and thus UV flux, are decreased \citep[e.g.][]{Yorke2008,Hosokawa2009,Hosokawa2010}. 

Ascertaining further how selection effects influence the difference in the distribution of RMS YSOs and UC\HII{} regions will require the luminosity functions of these sources, which will be the subject of a future paper (Mottram et al., 2010, in prep.).

\section{Conclusions}
\label{S:conclusions}

After preforming tests of the influence of the input parameters to the model SED fitter of \citet{Robitaille2007a}, we have obtained the luminosities of 1173 young RMS sources, of which 1069 have unique kinematic distances and good fits.

The luminosity distributions of these sources in terms of their RMS classification have been presented, and show that there are few MYSOs with L~$\geq$~10$^{5}$\lsol{}, though we detect UC\HII{} regions up to $\sim$7~$\times$~10$^{5}$\lsol{}, a difference which we find to be statistically significant. We also find that using the SED fitter to obtain bolometric flux measurements is more consistent and has less scatter, than either using simple trapezium rule integration or greybody fits for mid-IR bright young massive stars. Finally, we obtained the flux ratios consistent with our SED fits, which allow us to estimate the bolometric flux, of 280 sources for which far-IR fluxes could not be obtained.

Having obtained the luminosity distributions of YSOs and \HII{} regions in the RMS sample, we can go on to calculate the luminosity functions for these sources. This will be discussed in a forthcoming paper.

\begin{acknowledgements}

The authors would like to thank the referee, Barbara Whitney, for her helpful comments and suggestions which improved the quality and clarity of this paper. JCM is partially funded by a Postgraduate Studentship and by a Postdoctoral Research Associate grant from the Science and Technologies Research Council of the United Kingdom (STFC). This publication makes use of data products from the Two Micron All Sky Survey, which is a joint project of the University of Massachusetts and the Infrared Processing and Analysis Center/California Institute of Technology, funded by the National Aeronautics and Space Administration and the National Science Foundation. This paper made use of information from the Red MSX Source survey database at www.ast.leeds.ac.uk/RMS which was constructed with support from the Science and Technology Facilities Council of the UK.

\end{acknowledgements}

\label{lastpage}


\begin{thebibliography}{52}
\expandafter\ifx\csname natexlab\endcsname\relax\def\natexlab#1{#1}\fi

\bibitem[{{Beltr{\'a}n} {et~al.}(2006){Beltr{\'a}n}, {Brand}, {Cesaroni},
  {Fontani}, {Pezzuto}, {Testi}, \& {Molinari}}]{Beltran2006}
{Beltr{\'a}n}, M.~T., {Brand}, J., {Cesaroni}, R., {et~al.} 2006, \aap, 447,
  221

\bibitem[{{Benjamin} {et~al.}(2003){Benjamin}, {Churchwell}, {Babler}, {Bania},
  {Clemens}, {Cohen}, {Dickey}, {Indebetouw}, {Jackson}, {Kobulnicky},
  {Lazarian}, {Marston}, {Mathis}, {Meade}, {Seager}, {Stolovy}, {Watson},
  {Whitney}, {Wolff}, \& {Wolfire}}]{Benjamin2003}
{Benjamin}, R.~A., {Churchwell}, E., {Babler}, B.~L., {et~al.} 2003, \pasp,
  115, 953

\bibitem[{{Beuther} {et~al.}(2002){Beuther}, {Schilke}, {Menten}, {Motte},
  {Sridharan}, \& {Wyrowski}}]{Beuther2002a}
{Beuther}, H., {Schilke}, P., {Menten}, K.~M., {et~al.} 2002, \apj, 566, 945

\bibitem[{{Brand} \& {Blitz}(1993)}]{Brand1993}
{Brand}, J. \& {Blitz}, L. 1993, \aap, 275, 67

\bibitem[{{Cao} {et~al.}(1997){Cao}, {Terebey}, {Prince}, \&
  {Beichman}}]{Cao1997}
{Cao}, Y., {Terebey}, S., {Prince}, T.~A., \& {Beichman}, C.~A. 1997, \apjs,
  111, 387

\bibitem[{{Carey} {et~al.}(2009){Carey}, {Noriega-Crespo}, {Mizuno}, {Shenoy},
  {Paladini}, {Kraemer}, {Price}, {Flagey}, {Ryan}, {Ingalls}, {Kuchar},
  {Pinheiro Gon{\c c}alves}, {Indebetouw}, {Billot}, {Marleau}, {Padgett},
  {Rebull}, {Bressert}, {Ali}, {Molinari}, {Martin}, {Berriman}, {Boulanger},
  {Latter}, {Miville-Deschenes}, {Shipman}, \& {Testi}}]{Carey2009}
{Carey}, S.~J., {Noriega-Crespo}, A., {Mizuno}, D.~R., {et~al.} 2009, \pasp,
  121, 76

\bibitem[{{Churchwell} {et~al.}(2009){Churchwell}, {Babler}, {Meade},
  {Whitney}, {Benjamin}, {Indebetouw}, {Cyganowski}, {Robitaille}, {Povich},
  {Watson}, \& {Bracker}}]{Churchwell2009}
{Churchwell}, E., {Babler}, B.~L., {Meade}, M.~R., {et~al.} 2009, \pasp, 121,
  213

\bibitem[{{Clarke} {et~al.}(2006){Clarke}, {Lumsden}, {Oudmaijer}, {Busfield},
  {Hoare}, {Moore}, {Sheret}, \& {Urquhart}}]{Clarke2006}
{Clarke}, A.~J., {Lumsden}, S.~L., {Oudmaijer}, R.~D., {et~al.} 2006, \aap,
  457, 183

\bibitem[{{Cohen} {et~al.}(2000){Cohen}, {Hammersley}, \& {Egan}}]{Cohen2000}
{Cohen}, M., {Hammersley}, P.~L., \& {Egan}, M.~P. 2000, \aj, 120, 3362

\bibitem[{{de Wit} {et~al.}(2010){de Wit}, {Hoare}, {Oudmaijer}, \&
  {Lumsden}}]{deWit2010}
{de Wit}, W.~J., {Hoare}, M.~G., {Oudmaijer}, R.~D., \& {Lumsden}, S.~L. 2010,
  \aap, 515, A45

\bibitem[{{Di Francesco} {et~al.}(2008){Di Francesco}, {Johnstone}, {Kirk},
  {MacKenzie}, \& {Ledwosinska}}]{DiFrancesco2008}
{Di Francesco}, J., {Johnstone}, D., {Kirk}, H., {MacKenzie}, T., \&
  {Ledwosinska}, E. 2008, \apjs, 175, 277

\bibitem[{{Egan} {et~al.}(2003){Egan}, {Price}, {Kraemer}, {Mizuno}, {Carey},
  {Wright}, {Engelke}, {Cohen}, \& {Gugliotti}}]{Egan2003a}
{Egan}, M.~P., {Price}, S.~D., {Kraemer}, K.~E., {et~al.} 2003, VizieR Online
  Data Catalog, 5114, 0

\bibitem[{{Egan} {et~al.}(1999){Egan}, {Price}, {Moshir}, {Cohen}, \&
  {Tedesco}}]{Egan1999}
{Egan}, M.~P., {Price}, S.~D., {Moshir}, M.~M., {Cohen}, M., \& {Tedesco}, E.
  1999, NASA STI/Recon Technical Report, 14854

\bibitem[{{Emerson}(1988)}]{Emerson1988}
{Emerson}, J.~P. 1988, in NATO ASIC Proc. 241: Formation and Evolution of Low
  Mass Stars, ed. A.~K. {Dupree} \& M.~T.~V.~T. {Lago}, 193

\bibitem[{{Fa{\'u}ndez} {et~al.}(2004){Fa{\'u}ndez}, {Bronfman}, {Garay},
  {Chini}, {Nyman}, \& {May}}]{Faundez2004}
{Fa{\'u}ndez}, S., {Bronfman}, L., {Garay}, G., {et~al.} 2004, \aap, 426, 97

\bibitem[{{Hanson} {et~al.}(2002){Hanson}, {Luhman}, \& {Rieke}}]{Hanson2002}
{Hanson}, M.~M., {Luhman}, K.~L., \& {Rieke}, G.~H. 2002, \apjs, 138, 35

\bibitem[{{Hill} {et~al.}(2005){Hill}, {Burton}, {Minier}, {Thompson}, {Walsh},
  {Hunt-Cunningham}, \& {Garay}}]{Hill2005}
{Hill}, T., {Burton}, M.~G., {Minier}, V., {et~al.} 2005, \mnras, 363, 405

\bibitem[{{Hill} {et~al.}(2009){Hill}, {Pinte}, {Minier}, {Burton}, \&
  {Cunningham}}]{Hill2009}
{Hill}, T., {Pinte}, C., {Minier}, V., {Burton}, M.~G., \& {Cunningham}, M.~R.
  2009, \mnras, 392, 768

\bibitem[{{Hoare} \& {Franco}(2007)}]{Hoare2007b}
{Hoare}, M.~G. \& {Franco}, J. 2007, in Diffuse Matter from Star Forming
  Regions to Active Galaxies, ed. {Hartquist, T.~W., Pittard, J.~M., \& Falle,
  S.~A.~E.~G.}, 61

\bibitem[{{Hoare} {et~al.}(2005){Hoare}, {Lumsden}, {Oudmaijer}, {Urquhart},
  {Busfield}, {Sheret}, {Clarke}, {Moore}, {Allsopp}, {Burton}, {Purcell},
  {Jiang}, \& {Wang}}]{Hoare2005}
{Hoare}, M.~G., {Lumsden}, S.~L., {Oudmaijer}, R.~D., {et~al.} 2005, in IAU
  Symposium, Vol. 227, Massive Star Birth: A Crossroads of Astrophysics, ed.
  R.~{Cesaroni}, M.~{Felli}, E.~{Churchwell}, \& M.~{Walmsley}, 370--375

\bibitem[{{Hoare} {et~al.}(1991){Hoare}, {Roche}, \& {Glencross}}]{Hoare1991}
{Hoare}, M.~G., {Roche}, P.~F., \& {Glencross}, W.~M. 1991, \mnras, 251, 584

\bibitem[{{Hosokawa} \& {Omukai}(2009)}]{Hosokawa2009}
{Hosokawa}, T. \& {Omukai}, K. 2009, \apj, 691, 823

\bibitem[{{Hosokawa} {et~al.}(2010){Hosokawa}, {Yorke}, \&
  {Omukai}}]{Hosokawa2010}
{Hosokawa}, T., {Yorke}, H.~W., \& {Omukai}, K. 2010, \apj, 721, 478

\bibitem[{{Indebetouw} {et~al.}(2005){Indebetouw}, {Mathis}, {Babler}, {Meade},
  {Watson}, {Whitney}, {Wolff}, {Wolfire}, {Cohen}, {Bania}, {Benjamin},
  {Clemens}, {Dickey}, {Jackson}, {Kobulnicky}, {Marston}, {Mercer},
  {Stauffer}, {Stolovy}, \& {Churchwell}}]{Indebetouw2005}
{Indebetouw}, R., {Mathis}, J.~S., {Babler}, B.~L., {et~al.} 2005, \apj, 619,
  931

\bibitem[{{Jackson} {et~al.}(2006){Jackson}, {Rathborne}, {Shah}, {Simon},
  {Bania}, {Clemens}, {Chambers}, {Johnson}, {Dormody}, {Lavoie}, \&
  {Heyer}}]{Jackson2006}
{Jackson}, J.~M., {Rathborne}, J.~M., {Shah}, R.~Y., {et~al.} 2006, \apjs, 163,
  145

\bibitem[{{Keto}(2003)}]{Keto2003}
{Keto}, E. 2003, \apj, 599, 1196

\bibitem[{{Kim} {et~al.}(1994){Kim}, {Martin}, \& {Hendry}}]{Kim1994}
{Kim}, S.-H., {Martin}, P.~G., \& {Hendry}, P.~D. 1994, \apj, 422, 164

\bibitem[{{Liszt} {et~al.}(1981){Liszt}, {Burton}, \& {Bania}}]{Liszt1981}
{Liszt}, H.~S., {Burton}, W.~B., \& {Bania}, T.~M. 1981, \apj, 246, 74

\bibitem[{{Lumsden} {et~al.}(2002){Lumsden}, {Hoare}, {Oudmaijer}, \&
  {Richards}}]{Lumsden2002}
{Lumsden}, S.~L., {Hoare}, M.~G., {Oudmaijer}, R.~D., \& {Richards}, D. 2002,
  \mnras, 336, 621

\bibitem[{{Martins} {et~al.}(2005){Martins}, {Schaerer}, \&
  {Hillier}}]{Martins2005}
{Martins}, F., {Schaerer}, D., \& {Hillier}, D.~J. 2005, \aap, 436, 1049

\bibitem[{{McClure-Griffiths} {et~al.}(2005){McClure-Griffiths}, {Dickey},
  {Gaensler}, {Green}, {Haverkorn}, \& {Strasser}}]{McClure-Griffiths2005}
{McClure-Griffiths}, N.~M., {Dickey}, J.~M., {Gaensler}, B.~M., {et~al.} 2005,
  \apjs, 158, 178

\bibitem[{{Meynet} \& {Maeder}(2000)}]{Meynet2000}
{Meynet}, G. \& {Maeder}, A. 2000, \aap, 361, 101

\bibitem[{{Minier} {et~al.}(2005){Minier}, {Burton}, {Hill}, {Pestalozzi},
  {Purcell}, {Garay}, {Walsh}, \& {Longmore}}]{Minier2005}
{Minier}, V., {Burton}, M.~G., {Hill}, T., {et~al.} 2005, \aap, 429, 945

\bibitem[{{Mottram}(2008)}]{Mottram2008}
{Mottram}, J.~C. 2008, PhD thesis, University of Leeds (UK)

\bibitem[{{Mottram} {et~al.}(2010){Mottram}, {Hoare}, {Lumsden}, {Oudmaijer},
  {Urquhart}, {Meade}, {Moore}, \& {Stead}}]{Mottram2010}
{Mottram}, J.~C., {Hoare}, M.~G., {Lumsden}, S.~L., {et~al.} 2010, \aap, 510,
  A260000+

\bibitem[{{Mottram} {et~al.}(2007){Mottram}, {Hoare}, {Lumsden}, {Oudmaijer},
  {Urquhart}, {Sheret}, {Clarke}, \& {Allsopp}}]{Mottram2007}
{Mottram}, J.~C., {Hoare}, M.~G., {Lumsden}, S.~L., {et~al.} 2007, \aap, 476,
  1019

\bibitem[{{Rathborne} {et~al.}(2009){Rathborne}, {Johnson}, {Jackson}, {Shah},
  \& {Simon}}]{Rathborne2009}
{Rathborne}, J.~M., {Johnson}, A.~M., {Jackson}, J.~M., {Shah}, R.~Y., \&
  {Simon}, R. 2009, \apjs, 182, 131

\bibitem[{{Robitaille}(2008)}]{Robitaille2008}
{Robitaille}, T.~P. 2008, in Astronomical Society of the Pacific Conference
  Series, Vol. 387, Massive Star Formation: Observations Confront Theory, ed.
  H.~{Beuther}, H.~{Linz}, \& T.~{Henning}, 290

\bibitem[{{Robitaille} {et~al.}(2007){Robitaille}, {Whitney}, {Indebetouw}, \&
  {Wood}}]{Robitaille2007a}
{Robitaille}, T.~P., {Whitney}, B.~A., {Indebetouw}, R., \& {Wood}, K. 2007,
  \apjs, 169, 328

\bibitem[{{Robitaille} {et~al.}(2006){Robitaille}, {Whitney}, {Indebetouw},
  {Wood}, \& {Denzmore}}]{Robitaille2006}
{Robitaille}, T.~P., {Whitney}, B.~A., {Indebetouw}, R., {Wood}, K., \&
  {Denzmore}, P. 2006, \apjs, 167, 256

\bibitem[{{Roman-Duval} {et~al.}(2009){Roman-Duval}, {Jackson}, {Heyer},
  {Johnson}, {Rathborne}, {Shah}, \& {Simon}}]{RomanDuval2009}
{Roman-Duval}, J., {Jackson}, J.~M., {Heyer}, M., {et~al.} 2009, \apj, 699,
  1153

\bibitem[{{Skrutskie} {et~al.}(2006){Skrutskie}, {Cutri}, {Stiening},
  {Weinberg}, {Schneider}, {Carpenter}, {Beichman}, {Capps}, {Chester},
  {Elias}, {Huchra}, {Liebert}, {Lonsdale}, {Monet}, {Price}, {Seitzer},
  {Jarrett}, {Kirkpatrick}, {Gizis}, {Howard}, {Evans}, {Fowler}, {Fullmer},
  {Hurt}, {Light}, {Kopan}, {Marsh}, {McCallon}, {Tam}, {Van Dyk}, \&
  {Wheelock}}]{Skrutskie2006}
{Skrutskie}, M.~F., {Cutri}, R.~M., {Stiening}, R., {et~al.} 2006, \aj, 131,
  1163

\bibitem[{{Urquhart} {et~al.}(2007{\natexlab{a}}){Urquhart}, {Busfield},
  {Hoare}, {Lumsden}, {Clarke}, {Moore}, {Mottram}, \&
  {Oudmaijer}}]{Urquhart2007a}
{Urquhart}, J.~S., {Busfield}, A.~L., {Hoare}, M.~G., {et~al.}
  2007{\natexlab{a}}, \aap, 461, 11

\bibitem[{{Urquhart} {et~al.}(2007{\natexlab{b}}){Urquhart}, {Busfield},
  {Hoare}, {Lumsden}, {Oudmaijer}, {Moore}, {Gibb}, {Purcell}, {Burton}, \&
  {Marechal}}]{Urquhart2007c}
{Urquhart}, J.~S., {Busfield}, A.~L., {Hoare}, M.~G., {et~al.}
  2007{\natexlab{b}}, \aap, 474, 891

\bibitem[{{Urquhart} {et~al.}(2008{\natexlab{a}}){Urquhart}, {Busfield},
  {Hoare}, {Lumsden}, {Oudmaijer}, {Moore}, {Gibb}, {Purcell}, {Burton},
  {Mar{\'e}chal}, {Jiang}, \& {Wang}}]{Urquhart2008a}
{Urquhart}, J.~S., {Busfield}, A.~L., {Hoare}, M.~G., {et~al.}
  2008{\natexlab{a}}, \aap, 487, 253

\bibitem[{{Urquhart} {et~al.}(2008{\natexlab{b}}){Urquhart}, {Hoare},
  {Lumsden}, {Oudmaijer}, \& {Moore}}]{Urquhart2008b}
{Urquhart}, J.~S., {Hoare}, M.~G., {Lumsden}, S.~L., {Oudmaijer}, R.~D., \&
  {Moore}, T.~J.~T. 2008{\natexlab{b}}, in Astronomical Society of the Pacific
  Conference Series, Vol. 387, Massive Star Formation: Observations Confront
  Theory, ed. H.~{Beuther}, H.~{Linz}, \& T.~{Henning}, 381

\bibitem[{{Urquhart} {et~al.}(2009){Urquhart}, {Hoare}, {Purcell}, {Lumsden},
  {Oudmaijer}, {Moore}, {Busfield}, {Mottram}, \& {Davies}}]{Urquhart2009}
{Urquhart}, J.~S., {Hoare}, M.~G., {Purcell}, C.~R., {et~al.} 2009, \aap, 501,
  539

\bibitem[{{Urquhart} {et~al.}(2010){Urquhart}, {Moore}, {Hoare}, {Lumsden},
  {Oudmaijer}, {Rathborne}, {Mottram}, {Davies}, \& {Stead}}]{Urquhart2010}
{Urquhart}, J.~S., {Moore}, T.~J.~T., {Hoare}, M.~G., {et~al.} 2010, \mnras,
  501

\bibitem[{{Walmsley}(1995)}]{Walmsley1995}
{Walmsley}, M. 1995, in Revista Mexicana de Astronomia y Astrofisica Conference
  Series, Vol.~1, Revista Mexicana de Astronomia y Astrofisica Conference
  Series, ed. S.~{Lizano} \& J.~M. {Torrelles}, 137

\bibitem[{{Wood} \& {Churchwell}(1989)}]{Wood1989a}
{Wood}, D.~O.~S. \& {Churchwell}, E. 1989, \apjs, 69, 831

\bibitem[{{Yorke} \& {Bodenheimer}(2008)}]{Yorke2008}
{Yorke}, H.~W. \& {Bodenheimer}, P. 2008, in Astronomical Society of the
  Pacific Conference Series, Vol. 387, Massive Star Formation: Observations
  Confront Theory, ed. H.~{Beuther}, H.~{Linz}, \& T.~{Henning}, 189

\bibitem[{{Yorke} \& {Sonnhalter}(2002)}]{Yorke2002}
{Yorke}, H.~W. \& {Sonnhalter}, C. 2002, \apj, 569, 846

\end{thebibliography}
\end{document}